\pdfoutput=1
\documentclass[11pt]{article}
\usepackage{natbib}
\usepackage{times}
\usepackage{latexsym}
\usepackage[T1]{fontenc}
\usepackage[utf8]{inputenc}
\usepackage{microtype}
\usepackage{inconsolata}
\usepackage{graphicx}
\usepackage{datetime2}
\usepackage{footnote}
\usepackage{authblk}
\usepackage[hyphens]{url}
\usepackage{hyperref}

\title{Online Influence Campaigns: Strategies and Vulnerabilities}

\author[1,2,3]{Andreea Musulan}
\author[1,4]{Veronica Xia}
\author[1]{Ethan Kosak-Hine}
\author[1]{Tom Gibbs}
\author[1,4]{Vidya Sujaya}
\author[1,4]{Reihaneh Rabbany}
\author[1,2]{Jean-François Godbout\thanks{Equal advising.}}
\author[1,4]{Kellin Pelrine\textsuperscript{*}}
\affil[1]{Mila}
\affil[2]{Université de Montréal}
\affil[3]{IVADO}
\affil[4]{McGill University}

\begin{document}

\pagenumbering{arabic}

\maketitle

\begin{abstract}
In order to combat the creation and spread of harmful content online, this paper defines and contextualizes the concept of inauthentic, societal-scale manipulation by malicious actors. We review the literature on societally harmful content and how it proliferates to analyze the manipulation strategies used by such actors and the vulnerabilities they target. We also provide an overview of three case studies of extensive manipulation campaigns to emphasize the severity of the problem. We then address the role that Artificial Intelligence plays in the development and dissemination of harmful content, and how its evolution presents new threats to societal cohesion for countries across the globe. Our survey aims to increase our understanding of not just particular aspects of these threats, but also the strategies underlying their deployment, so we can effectively prepare for the evolving cybersecurity landscape.
\end{abstract}

\section{Introduction}

One of the greatest threats to social cohesion for countries worldwide stems from \textit{inauthentic, societal-scale manipulation} by \textit{malicious actors}. This concept broadly refers to the idea that actors, foreign or domestic, engage in efforts to influence the stability of societies by operating information-based influence and reconnaissance campaigns, while concealing their origins and harmful intentions. In this paper, we provide an overview of this concept by reviewing the current state of research on the topic and evaluating the most common mitigation strategies proposed by scholars, policymakers, and intelligence communities. 

The efforts of these malicious actors can be targeted towards public or private networks, online and offline. This review focuses on digital manifestations of societal-scale manipulation. Users on public networks --- such as social media platforms --- can be victims of deceptive campaigns deployed by malicious actors. These campaigns are meant to shape public opinion, usually during a particular event that bears consequences for a given society as a whole, such as an election. 

Individuals using private networks to exchange sensitive or confidential information --- such as those used by government, corporations, and financial institutions --- can also be victims of manipulation campaigns. The goal of malicious actors that target these networks is to obtain private information. Like public opinion manipulation on public networks, the infiltration of private networks also bears societal-scale consequences. The effects of these infiltrations go beyond compromising national security; they can lead members of the general public to question the capacity of, for example, political or financial institutions to safeguard their information. In order to maximize the effectiveness of these campaigns, malicious actors must undertake reconnaissance efforts on their targets. These actors collect and synthesize information to target specific individuals via social engineering, or develop user typologies based on the potential effectiveness of exposure to specific information and narratives. Thus, manipulation campaigns are highly pernicious to the stability of societies, as they threaten social cohesion by polarizing the general public and by compromising public trust in political institutions and leaders. 

While the origins of societal scale manipulation threats can differ, there are common vulnerabilities in online communities accessible to democratic citizens that can be exploited by malicious actors. It is now relatively easy to identify and target social media users who are more susceptible to manipulation by analyzing their digital footprint, including, but not limited to, their online behavior, engagement patterns, and personal profile information from social media platforms \citep{moti2024targeted, hermann2023psychological, marengo2023predicting, simchon2023online, chang2023cybersecurity, mchatton2023mitigating}. These vulnerabilities exist at multiple levels of abstraction, from the structural features of democratic societies to the social and psychological tendencies of individual public opinion formation. This type of targeted intervention is becoming even more sophisticated now with recent developments in generative Artificial Intelligence (AI), such as Large Language Models (LLMs). Through generative AI's audiovisual and text generation capabilities, this technology has significantly lowered the costs of producing convincing human-like content, which can be very difficult to detect in online communities \citep{de2023chatgpt, sun2023ai, cooke2024good, frank2024representative}. As a result, inauthentic, societal-scale manipulation is today considered to be one of the most important threats facing democracies worldwide \citep{EUAIAct2024, csis2023, bengio2024managing, wef2024}. 

The risks associated with such malicious activity explain why research on this topic has expanded significantly across various domains, including law \citep{susser2019online, spencer2020problem}, sociology \citep{krafft2020disinformation, schia2020hacking, giglietto2020takes, mathur2023manipulative}, political science \citep{rheault2021efficient, krafft2020disinformation, keller2019social}, communication \citep{rogers2020politics, walter2023trolls}, and engineering and computer science \citep{cresci2020decade, khaund2021social, ferrara2023social}. Global intelligence communities have also been actively monitoring malicious actors in order to prepare for the myriad of elections around the world this year \citep{mtac_2024sept17, CISA_2022, taiwanintelligence}. In fact, Russia, Iran, and China have all already been engaged in ``influence operations [aimed at] elections'' \citep[9]{microsoftdefensereport2024}. For the case studies in this review, we focus on the U.S. intelligence community's efforts to safeguard the U.S. presidential election in the fall of 2024, given its overall geopolitical importance.

In this work, we aim to answer two central research questions:

\begin{itemize}

    \item First, what are the strategies and goals of malicious actors behind online, inauthentic, societal-scale manipulation?

    \item Second, what are the most pressing types of threats posed by these actors? 

\end{itemize}

According to \citet{akhtar2024sok}, most research on malicious activity has focused on its specific components, such as the spread of inaccurate information, automated communication agents (i.e. social bots), and coordinated campaigns. Although detailed exploration of these components individually are essential to developing fine-grained understandings of manipulation on social media, meta analyses such as the one presented in this paper and work by \citet{akhtar2024sok} are also integral to illustrating the bigger picture of these threats. 

In contrast to the review by \citet{akhtar2024sok}, here we largely focus on coordinated societal-scale manipulation campaigns. As a result, we go into greater detail with respect to the types of information involved in these campaigns as well as their methods of deployment. We also contextualize coordinated malicious activity into a broader framework of societal-scale risks by connecting social media manipulation campaigns to traditional cybersecurity threats --- such as the infiltration of private networks through social engineering and phishing --- more explicitly. Furthermore, we provide additional details on the motivations involved in societal-level threats by identifying specific types of actors involved and follow up our discussion of these threats with case studies. 

In the following section, we more systematically define inauthentic, societal-scale manipulation by malicious actors. We then present an overview of the types of manipulative content and strategies, the vulnerabilities they target, and three case studies of malicious actor activities that can shape public discourse in the U.S. We identify two dimensions to malicious influence campaigns online. The first one relates to the type of information created and propagated, such as disinformation or malinformation. The second one focuses on the methods of deployment for the kinds of societally harmful content we address. The most pernicious vector of deployment for malicious campaigns involves the creation and operation of networks of inauthentic individuals online and news websites. These strategies are designed to target societal-, social-, and individual-level vulnerabilities that shape public opinion. Thus, after outlining the content and methods of deployment used by malicious actors, we turn towards explaining the vulnerabilities highlighted in academic research. Another major component of our review introduces case studies of manipulation. Our focus on the U.S. is based on the recent threats reported leading up to the 2024 presidential election. Finally, we follow up on these issues by outlining the risks posed by the potential applications of AI by malicious actors and the risks posed by AI evolution.

\section{Defining the Threat}

Inauthentic societal-scale manipulation involves the development and operation of largely digital information campaigns by malicious actors. In this section, we will define each of the components of this concept in order to frame our discussion of the threats posed by such manipulation campaigns. Outlining the types of malicious actors involved will also help illustrate the potential origins of these threats so that researchers and intelligence communities can develop effective countermeasures.

The definition of \textit{inauthentic} is adopted from emerging terminology in the social media space. In the context of this review, the term inauthentic is used to describe a collection of intricate and coordinated online activities undertaken by actors (individuals or groups) driven by the intent to deceive a target audience, such as voters during an election \citep{metainauthentic}. Unlike manipulation, which looks at the impact on people, inauthentic activity questions the motive of an actor. For example, as we will discuss at length below, on social media platforms such as Meta and X (formerly Twitter), inauthentic actors create fake accounts, deploy social bots, artificially boost content, and engage in trolling and spam, to mislead public discourse. Importantly, inauthentic behavior does not need to be false. Factual information can be presented in such a way as to be manipulative, but still be true. 

The influence of inauthentic actors ranges from individual decision-making to societal level public opinion formation. We specify \textit{societal-scale} to focus on high impact, widespread disruption; we consider behaviors that are a threat not only to individuals, but to the broader society. For instance, the current ``infodemic'' (misinformation epidemic) impacted the public health through COVID misinformation that spread online and through social media, which facilitated mistrust in political institutions \citep{gallotti2020assessing, eysenbach2020fight, yang2021covid, latkin2023assessment}. The term ``infodemic'' refers to the rapid and far-reaching spread of information of questionable quality \citep[1285]{gallotti2020assessing}. This phenomenon not only influenced individual lives, but also societal institutions such as hospitals and government policies. Another example of a threat that can pose societal-scale risks is the targeting of sensitive private networks through phishing, which involves attempts to collect confidential information by compromising user credentials \citep{microsoftthreatactors2024}. These attempts can have wide-reaching consequences if government,\footnote{For example, the phishing attacks on John Podesta led to a major controversy, as private information on Hilary Clinton's campaign manager was released \citep{bossetta2018weaponization}.} corporations,\footnote{For example, Microsoft routinely faces millions of phishing and other attacks due to how central it is to the ``digital ecosystem'' \citep[6]{microsoftdefensereport2024}.} or financial institutions\footnote{For example, the Credit Union Desjardins recently suffered a data breach that affected about 9.7 million of its customers in Canada and abroad \citep{dejardin2024}.} are the targets, because these networks contain centralized information on many individuals, or host information that can severely impact broader public opinion (e.g., political controversies) \citep{bossetta2018weaponization, priestman2019phishing, ali2024phishing}. 

\textit{Manipulation} refers to ``a kind of influence---an attempt to change the way someone would behave absent [of such] interventions'' \citep[13]{susser2019online}. It is an attempt to affect individuals' ``decision-making process[es]'' \citep[17]{susser2019online}. Manipulation can be more clearly defined when compared to other related concepts, such as coercion or persuasion. Coercion constrains available courses of actions for a given individual through an impending threat of consequence \citep{susser2019online}. Persuasion involves providing an individual with reasoning and/or information that would lead them to adopt a perspective or engage in an action \citep{susser2019online}. While both coercion and persuasion give individuals the opportunity to make an informed choice, manipulation often involves deception --- leading an individual to ``hold false beliefs'' \citep[21]{susser2019online} --- or other tactics that shape decision-making processes without the target's awareness \citep{susser2019online}.\footnote{For instance, another tool available to manipulators is distracting targets with extraneous information \citep[22]{susser2019online}.} A recent example of manipulation can be found in Russian-affiliated actors' use of fake news websites to influence public opinion of the U.S. government's support for Ukraine \citep{doppelganger}.

Finally, \textit{malicious actors} refer to individuals or groups, foreign or domestic, that operate information-based manipulation campaigns focused on shaping the opinions and actions of specific individuals in a way that advances ``a group or a nation's interests and objectives'' \citep{microsoftthreatactors2024}. To act maliciously is to engage in actions that are driven by the intent to cause harm to ``other people'' \citep{malicious}.\footnote{For example, during the 2016 U.S. presidential election, the Russian Internet Research Agency's focus on denigrating Hillary Clinton was intended to affect her chances of winning the election \citep{senaterussianreport}.} The types of actors involved in ``manipulating public opinion online'' \citep[1,9]{bradshaw2021industrialized} include: government-affiliated entities, political parties, private companies, members of civil society (citizens and organizations), and individuals motivated by financial gains \citep{bradshaw2021industrialized, microsoftthreatactors2024}.\footnote{For live monitoring of manipulation campaigns, it is useful to distinguish between identifiable actors and emerging threats whose affiliations have yet to be determined. The Microsoft Threat Analysis Center (MTAC) refers to these actors as ``groups in development,'' as their activities evolve \citep{microsoftthreatactors2024}.} 

Government entities refer here to ``communication or digital ministries, military campaigns, [...] police force activity [... and] state-funded media" \citep[8]{bradshaw2021industrialized}.\footnote{Government entities have not only been relying on the same resources used by criminals, but have also been known to work with them directly \citep[17]{microsoftdefensereport2024, sailio2020cyber}. These government actors can also be financially motivated \citep[17]{microsoftdefensereport2024}.} Often these entities have longstanding responsibilities for public opinion manipulation that go beyond temporary events such as elections and their general goal is to ``shape public attitudes'' \citep[8,17]{bradshaw2021industrialized}. Political parties tend to run these campaigns during election periods \citep[9]{bradshaw2021industrialized}. ``Nation-state'' actors such as governments and political parties focus ``traditional espionage or surveillance objectives'' on foreign government-affiliated actors, ``nongovernmental organizations'' (NGOs), and ``think tanks'' \citep{microsoftthreatactors2024}.\footnote{\citet[17]{bradshaw2021industrialized} specify three levels of capacity that are useful for categorizing the efforts of ``government or political party actors'' across countries: high, medium, and low capacity \citep[1,17]{bradshaw2021industrialized}. Capacity refers to the `` the number of government actors involved, the sophistication of tools, the number of campaigns, the size and permanency of teams, and [...] expenditures'' of the groups involved in manipulation \citep[17,18]{bradshaw2021industrialized}. At the most active level, large, longstanding teams engage in manipulation on an ongoing basis (including and beyond elections), may conduct research to improve their approaches, have designated funding (especially for ``state-sponsored media''), and operate domestic and foreign campaigns \citep[18]{bradshaw2021industrialized}. Medium capacity actors focus more on domestic manipulation but also have dedicated funding and talent to consistently influence public opinion \citep[18]{bradshaw2021industrialized}. Low capacity groups focus their efforts on specific events, such as elections, and operate exclusively at the domestic level \citep[18]{bradshaw2021industrialized}. } 

Another example of entities engaging in online manipulation campaigns are private companies that are contracted for developing ``computational propaganda'' and spreading ``disinformation for profit'' \citep[9]{bradshaw2021industrialized}.\footnote{The authors note that from 2009 to 2020, for the companies analyzed in their report, they found contracts for political campaigns added up to about \$60 million USD, although this is likely an underestimation of the true figure \citep[9]{bradshaw2021industrialized}.} These companies can develop ``cyberweapons'' for a market of consumers that then use those weapons for their own ends \citep{microsoftthreatactors2024}. These weapons have been used against ``global human rights efforts,'' to attack ``dissidents, human rights defenders, journalists, civil society advocates, and other private citizens'' \citep{microsoftthreatactors2024}. 

Individual ``citizen influencers'' and ``civil society'' groups\footnote{Which involve, more specifically ``civil society organizations, Internet subcultures, youth groups, hacker collectives, fringe movements, social media influencers, and volunteers''\citep[9]{bradshaw2021industrialized}.} that ``ideologically support a cause'' also represent another category of manipulative actors online \citep[9]{bradshaw2021industrialized}. These often work with other entities, whether the collaboration is ``implicitly and explicitly sanctioned by the state'' \citep[9]{bradshaw2021industrialized}.\footnote{\citet[9]{bradshaw2021industrialized} focus on explicit coordination in their report.} These actors can operate domestically, from foreign soil, or collaborate in the context of an international network. Political extremists, hacktivists, and terrorist groups would fall into this category \citep{CISA_2022, sailio2020cyber}.\footnote{Hacktivists are ideologically motivated ``activists'' that violate security policies using their computer hacking skills \citep[9]{sailio2020cyber}. Terrorist groups use a combination of coercive, manipulative, and persuasive tactics to advance their ideologically motivated goals \citep{sailio2020cyber, alonso2024coercion}.}

The final type of malicious actor is largely focused on financial rewards from spreading societally harmful content.\footnote{There are also actors that are not motivated by ideology or financial gain and do not necessarily coordinate their activities \citep{sailio2020cyber}. These actors may not always be malicious but compromise cyber security infrastructure for its own sake and can serve as ``cheap testing engineers'' if they engage in these activities as part of a ``well-organized vulnerability bounty program'' \citep[9]{sailio2020cyber}.} These actors can be ``criminal'' in nature if they, for example, engage in monetary ``extortion'' by making use of illegally obtained private information (i.e., through ``phishing'') \citep{microsoftthreatactors2024}.\footnote{Phishing is a method of obtaining private information from people online, where an attacker presents a link to a website that appears to be from a legitimate organization or individual but is actually malicious \citep{alkhalil2021phishing}. This concept will be defined in greater detail in the following section, specifically, the subsection on methods of deployment. On a related note, Microsoft has observed almost a threefold increase in the use of ``ransomware'' since last year \citep[27]{microsoftdefensereport2024}. Ransomware refers to harmful programs that restrict a user's ability to use the data on their devices until a payment is made to the attacker \citep{ransomware2020}.} However, these pecuniary-motivated individuals may not be criminal, as they can exploit features of specific platforms in order to achieve their goals. For example, on X, posts from users that pay for a ``blue-check'' (which was originally used to verify the identity of users) are given exposure priority on the platform and are able to receive ``a share of revenue from the ads in their replies,'' regardless of the veracity of the content they share \citep{bbcconspiracyprofit}. While individually, the behavior of these types of actors may not have societal-scale effects, through coordinated and/or simultaneous activity based on a common objective, they may influence global public opinion on a given issue \citep{najafabadi2018hacktivism}. 

The kind of manipulation we have described has far-reaching consequences that can extend beyond a given nation-state. Domestic politics, especially in powerful countries such as the U.S., can shape the broader geopolitical environment. Global perception of how control is exercised in these countries can influence whether and how, for example, trade and other partnerships develop between countries \citep{mansfield2017democracies, chow2017entry, bayer2004effects}. The perceived effectiveness of this manipulation can further serve as a signal to other threat actors that a given country is a vulnerable target \citep{tomz2007domestic}. Public disclosure of how manipulation campaigns are detected can also contribute to the evolution of content and strategies used by malicious actors, as these actors reflect on and adapt their approaches to make their operations increasingly covert. Our definition of \textit{inauthentic, societal-scale manipulation} by \textit{malicious actors} above helps set the stage for discussing the current state --- namely the content and strategies --- of its evolution today. 

\section{Manipulation Strategies}

What constitutes a manipulation strategy? As mentioned, there are two dimensions to the manipulation strategies implemented by these campaigns: types of content and methods of deployment. First, content refers to the type of information generated and spread to fulfill the manipulative objectives.\footnote{\citet{bradshaw2021industrialized} identify four categories of content types from a political perspective: ``pro-government or pro-party propaganda,'' political opposition attacks and ``smear campaigns,'' the use of ``trolling or harrassment'' to discourage ``political dissent and freedom of the press,'' and finally, stoking societal ``division'' and polarization \citep[13]{bradshaw2021industrialized}.} Misinformation, disinformation, malinformation, fake news, deepfakes, conspiracies, propaganda, rumors, and toxic speech, although distinct, all refer to the broader concept of societally harmful content, which is defined in greater detail below.\footnote{This list, while not exhaustive, covers the main forms of societally harmful content.} This kind of content can be malicious, and otherwise results in harmful effects on the receivers of the information. It is important to note that many of the forms of societally harmful content can be assisted or entirely generated by AI, such as LLMs and other generative AI tools designed to produce audiovisual content. 

Second, researchers have identified a variety of methods or strategies of deployment for societally harmful content. However, each of the methods can be facilitated by the creation or procurement of fictitious individuals and websites, either independently or --- more perniciously --- in entire coordinated groups \citep[1]{CISA_2022}. These are primarily used in interactions with individuals online, to share information, and support its perceived authenticity. ``Inauthentic personas'' can create accounts\footnote{These accounts can be used to maximize the effectiveness of their outreach efforts by taking advantage of the design of specific social media platforms. For example, on Reddit, users can create spaces for communities using subreddits \citep{weld2024making}, or on Facebook, users can create groups \citep{lee2014understanding, de2019analysis}.} on multiple social media platforms, posing as real individuals or ``experts'' \cite[1,3]{CISA_2022}. Fake websites, especially for news-related content, can also be used in conjunction with inauthentic personas to spread information. Although AI can be used to create these personas and websites, it is currently most likely to be used for content generation \citep[4]{openai2024cyber}. As these strategies continue to develop, actively monitoring their effectiveness will be pertinent to safeguarding political institutions, which is of paramount importance to ensuring the integrity of elections.

\subsection{Content}

Harmful content can lead to seriously detrimental impacts on society, such as division and polarization, so it is crucial to identify it and mitigate its spread. However, a precondition to that is understanding its different forms. Thus, below we describe each of the concepts identified as societally harmful content in turn. 

\textit{Misinformation} is commonly understood as the unintentional spread of ``false or misleading information'' \citep[1094]{lazer2018science}. Information that is false or misleading is ``inaccurate'' because it lacks ``clear evidence and expert opinion'' \citep[136]{vraga2020defining}. Although unintentionally spread, the threat of misinformation comes from its potential impact on individuals' ``beliefs'' \citep[136]{vraga2020defining}. In other words, consumption of misinformation can lead individuals to form inaccurate and unverified beliefs if they do not set out to investigate the veracity of the information, which they often do not \citep{vraga2020defining, susser2019online}. Especially in the case when there is a lack of information available to verify a claim, individuals have a ``tendency to become overconfident'' in their conclusions, and ``fill in the blanks'' themselves using their own knowledge and experiences \citep[794]{kuklinski2000misinformation}. Content that incites fear and/or involves controversial ``breaking news'' is also more likely to be spread and perceived to be credible \citep[7]{mtac_2024apr17}, as these features of information play on the psychological tendencies of humans \citep{gallotti2020assessing,ecker2022psychological}. Even when individuals are informed that their beliefs were constructed on the basis of inaccurate information, the effect of their former beliefs on their understanding of the world has a tendency to persist (referred to as the ``continued influence effect'') \citep[15]{ecker2022psychological}. As the adoption of these beliefs spreads across a given society, misinformation becomes more likely to shape public opinion. Although misinformation lacks the intent to cause harm, its spread could still be considered manipulation because it is likely that at least some of the individuals propagating it are aware of its inaccuracy, meaning it involves deliberate deception and becomes disinformation. 

\textit{Disinformation} is a form of misinformation that is distinguishable from the latter on the basis of intent. It is created and spread deliberately, with the explicit goal to deceive, manipulate or influence public opinion \citep{guess2020misinformation,lazer2018science}. In other words, disinformation is meant ``to cause harm'' \citep[142]{broda2024misinformation}. One of the ways disinformation has been used is in the case where there is a dearth of available factual information. ``Information gaps'' (i.e., lack of information on a particular claim) have been exploited by malicious actors to propagate misleading and inaccurate information that serves their ends \citep[2]{CISA_2022}.\footnote{Sometimes information gaps exist because, government officials want to prevent the incitement of panic among the general public. For example, in 2015 when South Korea was facing an epidemic, the government and news media refrained from publicizing which hospitals were afflicted \citep[227]{seo2021amplifying}. Unfortunately this had the opposite intended effect (i.e., it ``fueled fear'') and due to the lack of information, citizens used content from social media to inform themselves \citep[227]{seo2021amplifying}.} Actors that intentionally disseminate information they know to be false are generally malicious, and the action of spreading that content is manipulative as it can lead individuals to ``hold false beliefs'' \citep[21]{susser2019online}. 

\textit{Malinformation} often refers to ``information that is true, often private or confidential, that is intentionally leaked to inflict actual harm'' \citep{mcmahon2023maligned}. This concept also more generally includes information that is ``presented in a distorted manner'' \citep[2]{tomassi2024mapping}. An example of malinformation may include making confidential information and credentials available beyond the private sphere \citep[59]{csis2023}. Another example would be exaggerating, for example, the severity of a problem, such as the proliferation of a disease \citep{kandel2020information}. According to our definition of manipulation, the spread of malinformation is one of the tools available to manipulation campaigns, as it constitutes a way to influence the decision-making processes of individuals. This kind of information can be used to distract the general public from attending to information more pertinent to their, for example, vote choice during an election.  

\textit{Fake news} is fictitious information presented in the form of news articles \citep{lazer2018science}. The producers of this kind of fictitious content pose as news organizations but are not subject to the ``editorial norms and processes for ensuring the accuracy and credibility of information'' that legitimate news media adhere to \citep[1094]{lazer2018science}. Fake news shared by individuals or organizations that are unaware that the content is ``false or misleading'' (because of a dearth of rigorous journalistic practices) falls into the category of misinformation, whereas its deliberate spread would qualify as disinformation \citep[1094]{lazer2018science}. Fake news is especially engaging due to its ``novelty and negativity'' \citep[23]{csis2023}, as it provokes emotions such as ``fear, disgust, and surprise'' that make the information especially compelling for humans to share \citep[1146]{vosoughi2018spread}.\footnote{Automated social media accounts are equally likely to spread both fake and legitimate news, while humans have a greater tendency to share the former \citep{vosoughi2018spread}.} When an individual shares novel content, it gives their audiences the impression that they have access to ``inside information'' \citep[23]{csis2023}. Negative content attracts public attention because people have a greater tendency to focus on ``potential losses,'' as opposed to ``gains,'' and the reputations of the individuals sharing this kind of content benefit because they are ``warning others of potential threats'' \citep[23]{csis2023}.

\textit{Deepfakes} are audio, visual (i.e., images or video), and/or textual content that are generated using AI or deep learning technologies \citep{kietzmann2020deepfakes, csis2023}.\footnote{The term was actually created by an unidentified Reddit user that combined ``deep learning'' with ``fake'' \citep[136]{kietzmann2020deepfakes}. The user created and spread deepfake videos of famous individuals in pornographic contexts, and made the code they wrote for developing the videos available publicly, which led ``an explosion of fake content'' \citep[136]{kietzmann2020deepfakes}.} Deepfakes often depict people in situations that never happened or engaging in actions they never took \citep{csis2023}.\footnote{According to a recent Canadian Security Intelligence Service (CSIS) report on the technological capabilities of deepfakes, video content is more convincing and engaging than ``fabricated images and text'' \citep[22]{csis2023}.} While celebrities were the initial focus of visual deepfakes, teenagers and women have increasingly become more common targets \citep{deepfakewomen2024, kietzmann2020deepfakes}. The large majority ``of all deepfakes are pornographic and almost exclusively target women'' \citep{deepfakewomen2024}. However, deepfakes have also been used in political contexts. For example, a deepfake robocall featuring the voice likeness of Joe Biden ``falsely suggested that voting in the primary would preclude voters from casting a ballot in November'' \citep{bidenrobocall2024}. Another example was a deepfake video of ``President Obama swearing during a public service announcement'' \citep[135]{kietzmann2020deepfakes}. Often, this and other kinds of harmful content discussed in this review are spread on social media platforms by inauthentic or automated accounts. For example, the China-affiliated malicious actor Spamouflage used AI to create inauthentic social media accounts that spread unconvincing AI video content in 2020 \citep[4]{CISA_2022}.\footnote{\citet[15]{bradshaw2021industrialized} refers to this strategy as ``disinformation or manipulated media'' (such as false news, manipulated images and other modalities).} This kind of content is inherently deceptive, and the concept of misinformation could only apply for those sharing the content that are unaware of its lack of authenticity, while the content is exclusively disinformation by its creators.\footnote{An additional type of misleading content related to deepfakes are cheapfakes, which refer to visual content that is not manipulated using AI technology, but altered or generated by ``image-editing software products (like Photoshop, PremierePro, ...)'' \citep[9]{la2022multimodal}. Cheapfakes can also involve ``intentional image captions alteration, and speeding/slowing videos'' \citep[9]{la2022multimodal}. Another type of content that relies on AI that is deceptive is ``AI slop,'' which, like deepfakes, are AI-generated but appear ``uncanny and bizarre'' in character \citep{aislop2024}. AI slop generates user engagement on social media platforms by how outlandish the content is (making it humorous), which is beneficial for these platforms \citep{aislop2024}.} 

\textit{Conspiracy theories} are similar to malinformation in the sense that some may prove to be true, or partially accurate \citep{guess2020misinformation}. However, most often, these theories are false and can be classified as a type of misinformation unless the individual sharing them does not believe them to be true, in which case they would be considered a form of disinformation. What is unique to conspiracies is that it often involves the perception that a covert set of influential people ``control ... some aspect of society'' \citep[10]{guess2020misinformation}. People tend to share conspiracies on social media and ``selective[ly] expos[e]'' themselves to information that reinforces their beliefs, leading them to form, for example, ``echo chambers'' on social media \citep[22]{guess2020misinformation}.\footnote{We discuss echo chambers in the section on vulnerabilities.} The effects of propagating these theories extend beyond people's understandings of the conspiracies themselves, shaping individuals' ``entire worldview[s]'' \citep[1]{CISA_2022}. For example, state-affiliated actors from China spread the conspiracy that COVID-19 initially came from the U.S. and that ``a member of the U.S. military'' was responsible for its presence in China \citep[5]{CISA_2022}. Belief in this conspiracy theory could shape an individual's perception of the U.S. overall, given the severity of the claim. 

\textit{Propaganda} refers to ``any communications that are intended to persuade people to support one political group over another'' \citep[11]{guess2020misinformation}. These communications can involve accurate information, such as news articles from legitimate organizations, but is meant to ``disparage opposing viewpoints'' and ``rally public support,'' which can facilitate political polarization \citep[3]{tucker2018social}. Automated accounts are frequently involved in spreading this kind of information online \citep{tucker2018social}. The spread of propaganda is likely meant ``to ensure that some traditional media news stories are viewed more than others'' \citep[5]{tucker2018social}. Propaganda is a manipulation strategy because, although it may not involve deception, it constitutes as a decision-making process intervention that plays on the political biases of individuals without revealing the intention of the individuals spreading the information. In other words, given that the intention of propaganda is to polarize public opinion and this intention is not explicitly embedded in the information, it is a manipulation strategy.   

\textit{Rumors} could be considered ``a particular form of misinformation'' in the sense that they involve ``an acceptance of information that is factually unsubstantiated'' \citep[242]{berinsky2017rumors}. According to \citet{berinsky2017rumors}, this type of misinformation is distinguishable based on two key attributes: they ``are statements that lack specific standards of evidence'' and ``acquire their power through widespread social transmission'' \citep[242, 243]{berinsky2017rumors}. Like conspiracy theories, rumors can be true but would fall under disinformation if they are false and those sharing them are aware of their inaccuracy. Rumors could also be considered malinformation if they end up being at least partially substantiated and were intended to cause harm. Given the scale at which a rumor can be spread through the internet and social media, the transmission-based characteristic of rumors is especially pernicious to societal cohesion \citep{berinsky2017rumors}.

Finally, \textit{toxicity} is loosely defined in terms of the potential negative impact that a statement could have on public participation in discussions on internet forums and social media platforms \citep{rieder2021fabrics}. Toxicity can also be described as hate speech, harassment, threats, or ``forms of speech [...] that have a [...] tendency [...] to distress, insult, or annoy participants in a discussion'' \citep[2]{rieder2021fabrics}. The personal attacks on Hillary Clinton during the 2016 U.S. presidential election fall into this category of societally harmful content, as they centered on speaking ill of the candidate \citep{senaterussianreport}. Toxicity can be manipulative as it could negatively impact public opinion about an electoral candidate with no real reference to their leadership capacities. 

Digital space is rife with these forms of manipulative content, and while each of them may have slightly different effects, they can all contribute to the destabilization of societies, especially democratic ones. However, without effective approaches to deployment, these kinds of harmful content would not be able to influence public opinion at the societal level. In the following subsection, we discuss a variety of methods of content deployment, as well as other approaches that target private networks, that can have societal-level consequences.

\subsection{Methods of Deployment}

In order to spread societally harmful content, malicious actors primarily rely on individual or networks of inauthentic personas and websites that can inspire confidence among ``target audience[s]'' with respect to the content shared from these sources \citep[1]{CISA_2022}. Networks of fake personas and websites\footnote{As opposed to uncoordinated, stand-alone digital users and content.} pose a substantial threat to societal cohesion, as their coordinated activities and connections not only make the information they propagate more convincing, they can ensure harmful content reaches a wider audience. These tools can be created automatically with programming, manually, or using compromised identities of real people (i.e., ``hacked, stolen, or impersonation accounts'') \citep[11]{bradshaw2021industrialized}. Many of the methods of deployment discussed in this review involve the use of social bots, and/or inauthentic accounts more broadly, that engage in a variety of strategic behaviors. These accounts are especially useful for amplifying specific messages or creating the impression that an issue is of greater concern than it would otherwise be with solely organic engagement \citep{bradshaw2021industrialized, CISA_2022}. Social bot techniques include, for example, excessive posting, early and frequent retweeting of emerging news, tagging and/or mentioning influential figures, ``hashtag hijacking'' (where inauthentic users adopt an opponent’s hashtags to spam or otherwise undermine them), and flagging opponent’s legitimate content in hopes of remove it \citep[4]{himelein2021bots}. 

On social media, user accounts representing or imitating the characteristics of real people can be entirely automated (i.e., social bots), partially automated (i.e., cyborgs), or operated by humans \citep{bradshaw2021industrialized,NATOstratcom2018}.\footnote{Multiple accounts can be operated by a single individual or entity. Inauthentic accounts that are operated by a single individual for manipulation purposes are referred to as sock puppets \citep[59]{cook2014twitter}.} Automated accounts bolster specific ``narratives'' and overwhelm others \citep[11]{bradshaw2021industrialized}. Human-operated accounts have become ``increasingly more common,'' and can involve ``real and fake'' personas; they may also be cyborgs, that is, use some automation \citep[11]{bradshaw2021industrialized}. Often, the human component of account operation can come from ``freelance'' account operators (``usually employing people from developing countries'') \citep[9]{NATOstratcom2018}. Their purpose is to communicate with other users ``by posting [...], or by private messaging individuals'' \citep[11]{bradshaw2021industrialized}. Accounts involving misappropriated identities are less prevalent but can also pose a threat to ``groups, pages, or channels'' by taking advantage of the network controlled by the targeted users \citep[11]{bradshaw2021industrialized}.

These accounts can have varying levels of sophistication based on the extent and quality of content they create and/or share, whether the identity of the individuals operating the accounts are confirmed (i.e., account ``verification''), and the ``age'' of the accounts \citep[8]{NATOstratcom2018}.\footnote{The more content associated with the account, the more expensive, and the accounts offered could be: barren, contain ``a profile picture,'' ``a profile picture with a few photos,'' or ``a profile picture, photos, and a range of posts'' \citep[8]{NATOstratcom2018}. Account verification and older accounts are less likely to be detected and moderated by social media platforms, but they are also more expensive \citep[8]{NATOstratcom2018}. Verification requires a phone number or email address \citep[8]{NATOstratcom2018}.} The basic activities of these accounts are available to all users of a given social media platform, ceteris paribus. For most social media platforms, all users can post, comment, like, share, privately message, and report content and other users. Although often not permitted by these platforms, users can additionally engage in activity exchanges by mentioning each other in posts, so they can like and share each others' content to maximize exposure \citep{NATOstratcom2018}.  Furthermore, there are communication features that users, malicious or not, can use to bolster content. For example, on X, users can pay for a blue checkmark on their profile, making their content more likely to be promoted by X's recommendation algorithm \citep{bbcconspiracyprofit}. As another example, on Reddit, businesses can pay for their posts to be promoted on the platform \citep{redditads2024}.

Inauthentic and malicious accounts can be used to execute a type of ``Distributed Denial-of-Service (DDoS) attack'' \citep[15]{NATOstratcom2018}.\footnote{There are also ``Denial-of-Service (DoS) attack[s]'' which are not distributed, that is, do not rely on ``multiple computers'' \citep[134040]{falowo2022threat}.} Although DDoS attacks typically refer to the act of overloading servers with requests, in the context of social media platforms, these users can abuse safeguards in place for moderation (i.e., reporting features or deliberately posting content that leads to the removal of content or user accounts) \citep[15]{NATOstratcom2018}. This strategy has also been referred to as ``DDoS 2.0'' \citep[15]{NATOstratcom2018}, which involves the ``mass-reporting of content or accounts,'' and is often used to target political or otherwise influential public figures to reduce their visibility \citep[15]{bradshaw2021industrialized}.

``Trolling, doxing, or online harassment'' can also be used to target not only ``political opponents, activists, or journalists on social media,'' but regular people in order to create a toxic environment \citep[15]{bradshaw2021industrialized}. Trolling refers to a lack of decorum among users in online communities, or toxic behavior that rejects a given community's norms and rules \citep{cheng2017anyone}. Users engaging in this type of behavior are referred to as ``trolls'' that act independently or coordinate their activities \citep[353]{zannettou2019let}. When they are ``state-sponsored,'' trolls' actions are usually coordinated and motivated by the interests of the nation-state the accounts are affiliated with \citep[353]{zannettou2019let}.\footnote{For example, during the 2016 U.S. presidential election, trolls from Russia were supportive of Donald Trump, while trolls from Iran were against the candidate \citep{zannettou2019let}.} Doxing involves publicly sharing an individual's sensitive private information online, ``often with the intent to humiliate, threaten, intimidate, or punish'' them \citep[199]{douglas2016doxing}. Online harrassment, or bullying, can be used to stifle public discourse online by targeting negative comments towards specific users \citep{cheng2017anyone}. These tactics can affect individuals' reputations and generate toxicity in online communities, making real people less likely to engage in public discourse \citep{rieder2021fabrics, avalle2024persistent}.

The appearance of legitimacy for inauthentic social media accounts can be aided by supporting personas and content outside of these platforms but still on the internet. This involves creating the perception that these personas and websites represent ``experts'' (for example, journalists and researchers) and real news organizations \citep[1]{CISA_2022}. For example, the Russian-affiliated Doppelganger campaigns involved establishing networks of online news articles whose origins imitated legitimate American news websites \citep{doppelganger}. If websites shared by the aforementioned social media accounts appear credible, real users are more likely to open the links, which not only exposes them to the harmful content, but can also make them vulnerable to ``malicious software'' operating through, for example, ``browser extensions or malware'' \citep[9]{NATOstratcom2018}.\footnote{According to an analysis of cyber attack events from 2005 to 2021, in 2021, ``data breaches''  were the most common incidents, followed by malware, phishing and DoS/DDoS disruptions \citep[134045]{falowo2022threat}. Data breaches are incidents where private information is obtained by malicious actors, which can result in ``significant reputational damage, financial losses, and might be detrimental to the long-term stability of the impacted organization'' \citep[134041]{falowo2022threat}.}

On a related note, inauthentic personas and websites can be used by malicious actors for phishing attacks, which extend beyond social media platforms into private networks. These attacks are part of the broader concept of ``social engineering'' which involves attempts to manipulate a ``victim's emotions, gullibility, charity, or trust'' to achieve some broader goal \citep[13]{alabdan2020phishing}.\footnote{According to Microsoft, some of the most common approaches of malicious actors intent on infiltrating private networks involves ``social engineering -- specifically email phishing, SMS phishing, and voice phishing -- identity compromise, and exploiting vulnerabilities in public facing applications or unpatched operating systems'' \citep[27]{microsoftdefensereport2024}.} Social engineering, especially if targeted at specific individuals (e.g., OpenAI or government employees), requires reconnaissance on the victims in order to effectively manipulate them \citep{openai2024cyber}.\footnote{Reconnaissance involves obtaining information on a given targeted individual or entity to identify the vulnerabilities ``that will sooner or later be exploited'' \citep[134040]{falowo2022threat}.} For phishing, the goal is to collect private information on a target or set of targets \citep{alkhalil2021phishing, openai2024cyber}.\footnote{If the attacker's targets are specifically selected, it is referred to as ``spear-phishing'' \citep[2]{alkhalil2021phishing}.} This goal is achieved by attackers sharing website links from seemingly credible personas and sources via social media, email, phone calls, or private messaging \citep{bossetta2018weaponization, alkhalil2021phishing}. Victims that are convinced of the authenticity of these messages will open these malicious links, which are designed to gather private information (such as ``confidential or sensitive credentials'') \citep[3]{alkhalil2021phishing}. These links can prompt individuals to enter private information that would subsequently be used to compromise their accounts and allow attackers access to a given internal network (e.g., hosted by governments, corporations, or financial institutions) \citep{alkhalil2021phishing}.\footnote{Malicious actors can also use deepfakes to simulate legitimate ``biometric data'' to obtain access to security systems that utilize, for example, ``facial recognition'' \citep[61]{csis2023}.} The information obtained from these breaches may not be used immediately, especially if the infiltration attempts are sophisticated and focused on longer-term exploitation \citep{bhardwaj2020phishing, alkhalil2021phishing}. 

One of the most notable influence tactics used by malicious social media actors is related to social engineering and phishing. This ``data-driven strateg[y]''  involves the mass profiling individuals based on their interests and behaviors online in order to maximize the effectiveness of their exposure to, for example, ``political advertisements,'' which often involve contracting private companies \citep[15]{bradshaw2021industrialized}. \citet{CISA_2022} refers to this strategy as ``spread[ing] targeted content'' \citep[2]{CISA_2022}.\footnote{The essentially unfettered collection of social media user data has led to ``systemic predatory business practices'' \citep[326]{mannix2022personal}. ``Data brokers'' that obtain, synthesize, and sell user data from social media companies enable targeted campaigns \citep[331]{mannix2022personal}.} This approach requires collecting information on potential targets that indicate ``their worldview and interests'' \citep[2]{CISA_2022}. For example, Cambridge Analytica profiled American voters using Facebook data and targeted political messaging tailored to users based on their characteristics \citep{isaak2018user}. 

Another prominent manipulation tactic on social media involves ``flooding the information environment,'' as well as ``astroturfing'' \citep[1]{CISA_2022}. Flooding the information environment refers to producing and sharing content (i.e., ``posts and comment[s]'') with specific themes or ``narrative[s]'' that inundate public discourse \citep[1]{CISA_2022}. For example, while X users can incorporate hashtags into their content to reach a wider audience, there are no restrictions on the relevance of hashtag attribution to content \citep{najafabadi2018hacktivism}. This feature of X can allow users to attribute alternative meanings to the initial topic or issue associated with a given hashtag, also referred to as ``spam'' \citep{najafabadi2018hacktivism}.\footnote{Spamming is another category of manipulation strategies that trolling, harassment, and phishing fall into \citep[383]{yurtseven2021review}. Spamming can be used by malicious actors to target large groups of people with unsolicited, repetitive content \citep{yurtseven2021review}. Given that social media facilitates the ``massive broadcast'' of content, spammers often use the medium to spread their messages \citep[383]{yurtseven2021review}.} For example, on X, ``\#IranTalks'' originally referred to a set of ``tense nuclear negotiations'' from 2013 between Iran and several European countries \citep{najafabadi2018hacktivism}. \citet{najafabadi2018hacktivism} found that a set of coordinated X accounts were flooding the hashtag with unrelated content. Astroturfing also involves this kind of inundation, but content is framed to appear as organic ``support or opposition to a message, while concealing its true origin'' \citep[1]{CISA_2022}.\footnote{Astroturfing and flooding the information environment can lead individuals to restrain from sharing their opinions online, as they are deterred from contradicting the ``perceived majority'' \citep[395]{ross2019social}. This tendency to avoid confrontation can result ``spiral of silence,'' where ``a vocal minority'' can shape ``social norm[s]'' despite the fact that many individuals are not aligned with the minority's perspective \citep[395]{ross2019social}.} For example, during the 2016 election, Russia-affiliated actors posed as left- and right-leaning ``activists'' on social media that spammed public discussions with provocative messaging and encouraged ``activists to attend events'' \citep[6]{CISA_2022}.

Malicious actors may also make use of ``alternative social media platforms'' to amplify the salience of particular narratives to specific audiences on these smaller, less monitored and less regulated forums \citep[2]{CISA_2022}. For example, terrorist groups such as ``ISIS have leveraged [alternative] platforms to spread malign content, recruit new followers, and coordinate activities'' \citep[7]{CISA_2022}. These efforts can be effective as these activities facilitated a growth ``in terrorism attacks in Europe between 2015 and 2016'' \citep[7]{CISA_2022}. 

Another strategy used by malicious actors involves exploiting a lack of information on a specific claim circulating online. If there is a dearth of information available through search engines on a specific topic, piece of news, or other kind of information, malicious actors can capitalize on these ``information gaps'' by creating information of their own and ``seeding the search term on social media to encourage people to look it up'' \citep[2]{CISA_2022}. Since there is a lack of accurate information available to contradict the fabricated claims, it can deceive audiences \citep[2]{CISA_2022}. For example, in 2015, Russia-affiliated actors manipulated the lack of information on ``a Syrian humanitarian organization'' to create the impression that the group was associated ``with terrorism'' \citep[8]{CISA_2022}. 

Manipulation campaigns may further take advantage of influential non-malicious actors (i.e., ``individuals and organizations'') that share inaccurate or misleading information from a threat actor without knowing its veracity \citep[2]{CISA_2022}. An example can again be drawn from the Russian influence campaign aimed at the 2016 U.S. presidential election. Malicious actors shared polarizing messaging and used real people to do it, albeit, unwittingly \citep[9]{CISA_2022}. 

The networks of malicious actors and content discussed at the beginning of this section thus serve as agents and tools that can strengthen manipulation campaigns. These campaigns can incorporate a variety of different methods of deployment outlined above, from DDoS 2.0 attacks to deceiving real, influential people in order to spread a claim widely. The tactics we have described are meant to sow discord among citizens and public distrust of political institutions and leaders \citep[53]{csis2023}. This overview thus highlights some of the most common vectors of manipulation facing nation-states around the world. In the next section, we draw on academic research to inform us about the types of vulnerabilities that malicious actors can exploit. 

\section{Structural and Individual Vulnerabilities}

Manipulation campaigns exploit vulnerabilities that exist at three, overlapping levels of abstraction for democratic societies: structural or institutional features, social norms and practices, and individual psychological traits. Starting with the first, we define institutions as ``systems of established and prevalent social rules that structure social interactions'' \citep[2]{hodgson2006institutions}. These ``social rules'' can be used to take advantage of, or manipulate, entire populations. For example, in democratic societies, free speech is often a protected right, or at the very least, an expectation of their citizens \citep{saxonhouse2005free, krotoszynski2003comparative, goldman2019free}. While this right allows for open public discourse, it can be abused by malicious actors intent on sowing discord by spreading societally harmful content. Another example of a structural vulnerability of societies in general is the design of social media platforms, which facilitate the near instantaneous and wide-reaching spread of information, regardless of its veracity. The algorithms developed by social media platforms that determine what information to present to users, and in what order to present it, have been shown to increase the spread of misinformation \citep{pathak2023understanding}. As previously discussed, social bots have been able to influence online discourse by taking advantage of this structural feature of social media platforms.

Second, social norms and practices shape how and what kind of information is internalized by individuals in a way that makes it easy for malicious actors to infiltrate communities. For example, people have a tendency to communicate and connect with others aligned with their political ideology. As a group, politically aligned individuals commonly reject others with differing viewpoints. This leads to the development of ``echo chambers,'' which has been shown to propagate societally harmful content \citep{treen2020online}.

Third, at the level of the individual, known psychological traits can be taken advantage of by malicious actors. For example, in addition to the fact that people have a tendency to connect with others that share their views, individuals also tend to believe information that confirms what they already know \citep{ruffo2023studying}. Furthermore, individuals are particularly vulnerable to misinformation on a psychological level, given that less ``cognitive effort'' is required to consume it and it is also ``more emotional'' in nature when compared to ``factual news'' \citep[1]{carrasco2022fingerprints}.

Although we have already mentioned some of the vulnerabilities in the sections above, in this section, we go into greater detail and provide examples from academic research that cover the structural, social, and psychological targets of manipulation campaigns.

\subsection{Examples from the Literature}

Multiple levels of vulnerabilities are often exploited simultaneously through manipulation campaigns. This overlap is a consequence of the fact that there is a feedback loop between institutions that structure social interactions, which in turn shape individual decision-making processes, that eventually affect the development of institutions. For example, \citet{gallotti2020assessing} conduct a case study of the COVID-19 infodemic as it unfolded on X in a wide variety of languages. The COVID-19 infodemic proliferated for two main reasons. First, the design of social media platforms --- meant to spread information widely and nearly instantaneously --- facilitated the reach of related manipulation campaigns \citep{gallotti2020assessing}. Second, ``psychological mechanisms, such as curbing anxiety by denying or minimizing the seriousness of the threat'' were also exploited by campaigns that facilitated the infodemic \citep[1285]{gallotti2020assessing}. 

Another example of research that covers multiple levels of vulnerabilities is a case study on a different platform. \citet[554]{del2016spreading} investigate how Facebook users consume information related to scientific and conspiracy news by mapping the ``cascades'' of these kinds of information. Cascades --- which refer to the way in which information spreads on social media through, for example, replies and re-posts --- are produced by the nature of how social media platforms facilitate the spread of information (i.e., a structural feature of communication in digital societies) \citep[556]{del2016spreading}. These authors find that while science news diffused more quickly, it did not generate greater interest over time, but the opposite was true for conspiracy news \citep{del2016spreading}.\footnote{On the other hand, fake news propagates exponentially at an early stage of creation and can lead to rapid, significant loss in diffusion \citep{aimeur2023fake}. Furthermore, false information spreads faster than real information because of the lower barriers doing so (the information is not verified or substantiated but is sensational or provocative) \citep{aimeur2023fake}. This difference between the spread of conspiracies and fake news in dissemination may be due to the greater ``novelty'' of the latter \citep[23]{csis2023}. Since humans have a tendency to share new information \citep{vosoughi2018spread}, it is likely that the familiarity and complexity of conspiracy narratives (i.e. that a group of powerful individuals control society) makes people less likely to share them as quickly as fake breaking news.} People often share misinformation because their attention is focused on factors other than accuracy --- and therefore they fail to implement a strongly held preference for accurate information sharing \citep{pennycook2021shifting}. For example, when text is accompanied by an image in a message, it ``receive[s] more attention and reach[es] a wider audience'' \citep[1]{newman2023misinformed}. The image can appeal to individuals' preference for minimizing ``cognitive effort'' when consuming information \citep[1]{carrasco2022fingerprints}. These preferences correspond to individual-level features of human behavior and psychology. Social media platforms also allow for paid content promotion (regardless of its veracity), and content that inspires emotions such as ``fear, disgust, and surprise'' is especially compelling for humans to share \citep[1146]{vosoughi2018spread}, facilitating the spread of harmful information.

Partially due to the comparatively closed structure of Facebook's platform (as opposed to the openness of, for example, X), communities may have a tendency to be more homogeneous \citep{del2016spreading}. This ``social homogeneity'' was the primary driver of content diffusion in a study by \citet[554, 558]{del2016spreading}, which led to the formation of ``echo chambers.'' Whether a claim was accepted by an individual is strongly influenced by social norms and individual belief system \citep{del2016spreading}. The concept of ``social homogeneity'' and echo chambers is closely linked to homophily. Homophily refers to the ubiquitous tendency for humans to be linked to others who share their traits, observed in almost all (offline and online) social networks \citep{treen2020online}. This tendency is again exacerbated by the structure of social media platforms; they are designed to connect people on the basis of common interests \citep{treen2020online}. Like social homogeneity, homophily leads to echo chambers, which can exacerbate polarization, particularly between skeptic and activist groups \citep{treen2020online}.

\citet{zhen2023social} also find that misinformation spreads due to norms and structures of social networks. In their review, \citet{zhen2023social} investigate how social network dynamics and strategic action in online communities shape the transmission of misinformation, focusing on how such information about refugees and COVID-19 spread on X. These authors identify two social mechanisms that facilitate the spread of misinformation online: ``networked social influence and strategic information manipulation'' \citep[1]{zhen2023social}. The first concept refers to the spread of information due to the norms and structures of social networks, while the second focuses on the intentional distribution of such information through, for example, fake news websites and social bots \citep{zhen2023social}. A variant of homophiliy is political homophily, which describes the tendency of people to share information with those sharing their political ideology \citep{zhen2023social}. Similar to the concept of echo chambers, these authors use the term ``cyberbalkanization'' to describe the tendency that members of online communities have to primarily interact with other members at the cost of interactions with non-members \citep[20]{zhen2023social}. This social tendency makes these communities especially vulnerable to manipulation through exposure to propaganda. 

In an interdisciplinary review of fake news, polarization, and automation research, \citet{ruffo2023studying} find that repeatedly exposing social media users to unsubstantiated information increases their tendency to be credulous. This exposure is a function of social media platform content recommendation algorithms, which is a structural feature of social media. These authors discuss how this tendency is a consequence of, for example, the idea that people conform to behaviors and ideas of the social group they belong to \citep{ruffo2023studying}. At the individual level, \citet{ruffo2023studying} explain that people suffer from a number of psychological biases, related to but not exclusively, attentional, confirmation, congruence, belief, emotion, as well as selective exposure and overconfidence.\footnote{Attentional bias refers to individual's tendency to consume information that is related to what is already on their minds \citep{ruffo2023studying}. Confirmation bias describes the case where an individual relies on information that confirms their pre-existing beliefs and rejects contradictory claims \citep{ruffo2023studying, del2017modeling}. Confirmation bias has also been referred to as selective exposure \citep{ruffo2023studying}. Congruence bias is a related term that involves the same behaviors as those mentioned above, where people tend to rely on information that aligns with their worldviews and do not often search for contradictory information \citep{ruffo2023studying}. Fake news articles that contains weak claims or evidence but conclude with statements that align with a reader's ``expectations'' can lead to belief bias \citep[8]{ruffo2023studying}.  Emotional bias occurs when a piece of information contains high ``emotional valence,'' causing individuals to overlook facts stated by a specific individual because it ``upset[s] them'' \citep[8]{ruffo2023studying}. Overconfidence occurs when individuals' self-perception contains exaggerations of their abilities or when people minimize the abilities of ``opponent[s], the difficulty of a task, or possible risks'' \citep[317]{johnson2011evolution}. These individuals are overconfident in their ability to discern the veracity of information and rely heavily on their own knowledge or understanding to make sense of the world.} All of these kinds of biases are related to the concept of ``schemas,'' which refer to the human tendency to connect new information to pre-existing knowledge, as we attempt to make sense of the world
\citep[333]{mannix2022personal}

Malicious actors are aware of and capitalize on these vulnerabilities to make their manipulation campaigns more effective. In the next section, we present a series of case studies of these campaigns largely conducted by foreign adversaries. These state-operated or state-sanctioned campaigns were largely targeted at the U.S., and most of the following discussion focuses on the threat they pose to the American public. Given the far-reaching impact of U.S. presidential elections, it is important to discuss the state of information warfare, as it will help inform other countries, democratic or otherwise, of potential threats on the horizon. 

\section{Case Studies of Malicious Actors}

Although manipulation campaigns can be applied beyond political contexts, elections are especially vulnerable to this type of threat. The U.S. intelligence community distinguishes between election interference and influence campaigns \citep{usdeptstate}. Both concepts focus on foreign actors, which can be foreign governments or, for example, private companies enlisted by governments \citep{mtac_2023nov8}. Interference refers to attempts by foreign actors to ``degrade or disrupt'' a country's capacity to conduct elections \citep{usdeptstate}. This involves ``activities targeted at the technical aspects of the election, including voter registration'' \citep[1]{mtac_2023nov8}. Election interference is a part of influence that is focused on the practical conduct of elections. Election influence involves activities meant to ``shape election outcomes or undermine democratic processes'' \citep{usdeptstate}. These activities can be explicit or concealed, and can be targeted at ``candidates, political parties, voters or their preferences, or political processes'' in order to ``directly or indirectly'' shape election outcomes and processes \citep[1]{mtac_2023nov8}. 

The U.S. intelligence community is largely focused on election influence, likely because it is much more straightforward to monitor interference, given the fact that clear and effective safeguards can be put in place to minimize the potential for successful attempts at interference \citep{usdeptstate, usic_insiderthreats}. For example, auditing the technical components ``before, during, and after an election,'' or ``chain of custody procedures'' that log information on who was handling sensitive election documents and when, can stifle interference attempts \citep[4]{usic_insiderthreats}. On the other hand, electoral influence campaigns, like Doppelganger (discussed below), are more common, complex, and difficult to mitigate given the resources, skills, and time required to detect them. 

In this section, we present three case studies of foreign influence campaigns with geopolitical implications. We begin by describing two major manipulation campaigns active during the 2016 U.S. presidential election. We then discuss a campaign operated out of Russia that was mainly directed at shaping American public opinion towards U.S. involvement in the war in Ukraine. Finally, we compare ongoing influence campaign approaches from Russia, Iran, and China targeted at the 2024 U.S. presidential election. 

\subsection{2016 U.S. Presidential Election}

One of the most well-documented examples of societal-scale manipulation revolves around the 2016 U.S. presidential election, where the integrity of the electoral process was threatened by both domestic and foreign entities \citep{isaak2018user, senaterussianreport}. This case confirms the scale, scope, and severity of societal-scale manipulation attempts \citep{senaterussianreport}. First, voters were targeted by domestic actors.\footnote{Cambridge Analytica itself was a foreign company based in the United Kingdom, but a domestic election campaign team (supporting Trump's candidacy) utilized its services \citep{cambridgelondon2018}. Thus, manipulation campaigns can involve actors from multiple countries. Another example is European companies that use data-driven products from Russia, which in turn relies on ``manual labor'' from Asian countries \citep[326, 327]{mannix2022personal}.} Using account activity and profile information from social media platforms (especially Facebook), the information technology firm, Cambridge Analytica, developed psychological profiles of U.S. citizens for targeted advertising as part of Donald Trump's campaign \citep{isaak2018user}. The firm's goal ``was to identify those who might be enticed to vote for their client or be discouraged to vote for their opponent'' \citep{isaak2018user}. 

Second, the American public was manipulated by foreign actors. The U.S. Senate's Select Committee on Intelligence conducted an investigation into foreign interference during the 2016 election and released a report in 2019 \citep{senaterussianreport}. According to the report, Russia's Internet Research Agency set up social media accounts across platforms such as Facebook and X long before the election \citep{senaterussianreport}. These accounts, regardless of whether they appeared to be politically left- or right-leaning, spoke negatively of Hillary Clinton and positively about Donald Trump \citep{senaterussianreport}. These malicious Russian actors closely observed the impact of their efforts using metrics like ``comments, likes, re-posts, [and] changes in audience size'' \citep[27]{senaterussianreport}. After the election, these inauthentic accounts continued to provoke division among the American digital public, this time by promoting ``anti-Trump sentiment'' among left-leaning users \citep{senaterussianreport}.  In addition to the Committee's findings on these malicious accounts, the report also notes that social bots accounted for a substantial amount of public discussion of politics on X specifically \citep{senaterussianreport}. Research from the Oxford Internet Institute that sought to quantify the potential impact of bots on public discourse, malicious or not, found that automated accounts ``reached positions of measurable influence'' and would have been capable of shaping public discourse related to the election \citep[10,11]{senaterussianreport}.

Although the degree to which these influence campaigns affected the vote choice of the American public remains contested, the size and sophistication of these actors' operations reveal their resourcefulness, and highlighted vulnerabilities in U.S. national security. 

\subsection{Doppelganger}

One of the most recent examples of an extensive and well-resourced political manipulation project largely targeted at the U.S. that extended beyond an electoral context was ``Doppelganger'' \citep{doppelganger}. ``Doppelganger'' refers to a malicious actor that engaged in a series of ``foreign malign influence'' operations, organized by a set of Russian government affiliated companies, that began in 2022 \citep[1]{doppelganger}. These campaigns were largely meant to ``reduce international support for Ukraine, bolster pro-Russian policies and interests, and influence voters in the U.S. and foreign elections'' \citep[2]{doppelganger}. The companies involved purchased domains\footnote{Domains were purchased using cryptocurrency to obscure the affiliations of the actors involved because blockchain ledgers ``do not identify the individuals involved in the transaction[s],'' only their wallet (which are comprised of ``26-to-35-character-long case-sensitive string of letters and numbers'') and transaction identifiers (``a complex series of numbers'') \citep[35]{doppelganger}. Although difficult, it may be possible to attribute the transactions to specific individuals using ``tools that are available to law enforcement'' \citep[35]{doppelganger}.} that very closely resembled official U.S. news websites, such as the Washington Post, and used social media to direct traffic to Russian propaganda \citep{doppelganger}.\footnote{This is referred to as ``cybersquatting'' \citep[1]{doppelganger}. Furthermore, while the complete link to the article would lead to the fake news, the domains were not typically indexed, meaning, for example, ``a visit to www.foxnews.cx would reveal a blank or error page, [but] a visit to www.foxnews.cx/world/US-Decided-to-Trade-Ukraine-for-Security.html would reveal the active cybersquatted website with an article and the re-routing links to the legitimate Fox News'' website \citep[17]{doppelganger}.} The campaigns paid for advertising on social media platforms (content of which was sometimes AI generated), set up inauthentic social media accounts pretending to be Americans ``or other non-Russian[s]'' to bolster public engagement with the fake articles through ``comments,'' and used social media influencers to further generate traffic towards the propaganda \citep[2,16]{doppelganger}. Social bots were also used to facilitate the spread of the information and track the progress of the ``targeted social engineering'' campaigns \citep[30]{doppelganger}. Real users' reactions to the propaganda were tracked over time, and were used to tailor advertisements and engagement by the fake accounts \citep{doppelganger}. The approach used in this campaign focused on developing content (including memes, images, and videos) that contained ``a minimum of fake news and a maximum of realistic information'' to ensure the ``effective[ness]'' of the assets deployed \citep[32]{doppelganger}. 

Although measures of the engagement impact of Doppelganger operations are not currently available, information on the project's objectives with respect to the extent of their reach is publicly accessible \citep{doppelganger}. For example, these malicious actors systematically targeted communities on multiple social media platforms (such as X, Facebook, and Telegram), either by embedding themselves within existing groups or creating new ones \citep[169,173]{doppelganger}. These communities were selected based on the geographical location of their users, which were based in Ukraine \citep[169]{doppelganger}. In one of the campaigns, Doppelganger agents were instructed to post ``50,000 comments in 20 regions of Ukraine per month,'' with a goal of 10 million monthly engagements \citep[173]{doppelganger}. Thus, at the very least, the extent of the Doppelganger campaigns' reach reflect Russia's level of commitment to influencing public opinion. 

\subsection{2024 U.S. Presidential Election}

In a series of five reports released over the last year, the Microsoft Threat Analysis Center (MTAC) identified three nation-states involved in foreign influence activities directed at the 2024 U.S. presidential election: Russia, Iran, and China \citep{mtac_2023nov8,mtac_2024apr17,mtac_2024aug9,mtac_2024sept17, mtac_2024oct23}. Each of these actors took a slightly different approach to creating and deploying malicious influence campaigns based on their national interests and capacities.  

Malicious Russian actors' objectives included reducing public and policy support for Ukraine in the U.S., as well as NATO, while also sowing division among the American public to divert their attention away from international affairs \citep{mtac_2023nov8,mtac_2024apr17}. MTAC anticipated these actors to maintain the propagation of  ``divisive political content, staged videos, and even AI-enhanced propaganda'' leading up to the November election \citep[1,2]{mtac_2024sept17}. In all three of the last presidential races, Russian campaigns have centered on negatively impacting the left-leaning contender over the course of the final three months leading up to the election \citep[1]{mtac_2024sept17}.  

There were a number of malicious actors associated with Russia that each have their own approaches, strategies, and capabilities to achieve these objectives, most of which were connected directly or indirectly with the Russian state, reflecting the national character of their influence operations \citep[2]{mtac_2024apr17}. MTAC referred to one of these actors as Storm-1516, whose approach typically involved disinformation that originates from a supposed ``whistleblower or citizen journalist on a purpose-built video channel, which was then covered by a seemingly unaffiliated network of managed or affiliated websites'' \citep[2]{mtac_2024apr17}. According to MTAC, this actor was the most influential, and shifted its focus from Ukraine to the U.S. as the presidential election began, using fabricated video content, characteristic of the group \citep[5]{mtac_2024aug9}. When Vice President Kamala Harris became the Democratic candidate, a series of actors --- Storm-1516, Doppelganger, and Storm-1679 ---  scrambled to reorient their operations to target Harris instead of President Joe Biden \citep[1]{mtac_2024sept17,mtac_2024oct23}. Among Storm-1516's most recent activities, the actor created and spread fake videos centering on Harris' campaign \citep[2]{mtac_2024sept17,mtac_2024oct23}. Although Russian actors had increasingly made use of AI to generate content about Harris \citep{mtac_2024oct23}, they also relied on more basic approaches ``such as deceptive editing, spoofs, and staged videos'' \citep[4]{mtac_2024oct23}.\footnote{Spoofs are pieces of information (audio, visual, and text) that have been manipulated ``to misrepresent the identity of [its] sources and/or the validity and reliability'' of the content \citep[242]{innes2021disinformation}.} One of the videos asserted that in 2011 the candidate unintentionally hit and ``paralyzed'' someone with her car \citep[2]{mtac_2024sept17}. The video was originally shared through a fake news website purporting to be a local newspaper from San Francisco that established a presence online ``only [...] days beforehand'' \citep[2]{mtac_2024sept17}. Storm-1679 was also responsible for recently creating and spreading video content on X involving Harris, ``spoof[ing] news'' media and other entities such as ``Fox News, the FBI, and the technology publication Wired'' \citep[5]{mtac_2024oct23}.

MTAC mentioned that the use of ``cyber proxies and their amplifiers'' had become a newly incorporated strategy for Russian influence campaigns \citep[1]{mtac_2024sept17}.\footnote{A proxy server serves as an intermediary between an internet user and the websites they visit \citep{proxynordvpn2023}. Proxies conceal the actual internet protocol (IP) address of users, as they are IP addresses linked to other devices \citep{proxynordvpn2023}. They do not encrypt online activity, and are only used for individual applications, such as web browsers \citep{proxynordvpn2023}.} These proxies help malicious actors conceal their online identities while they collect data from ``hack-and-leak operation[s]'' (i.e., spreading malinformation and doxing) \citep[4]{mtac_2024sept17}. MTAC noted that these proxies could have been used to generate concern about disorder surrounding the election \citep[4]{mtac_2024sept17}. As an example of an electoral interference attempt, ``proxies and hacktivist groups'' allegedly attacked election websites related to the U.S. midterms in 2022 and the European Parliamentary Elections in 2024 \citep[3]{mtac_2024sept17}. 

Iran's attempts to influence the 2024 U.S. election were motivated by the war in Israel, which serves as a bulwark for the former's interest in stoking geopolitical tensions in the Middle East \citep[2]{mtac_2023nov8}. For the 2020 U.S. election, Iranian malicious actors that engaged in influence activities ``impersonated [...] extremists, […] attempted to sow discord among […] voters, […] [and] incite[d] violence against […] government officials'' \citep[2]{mtac_2023nov8} in order to generate confusion, emphasize ``existing political and social divisions,''\footnote{For example, along the lines of ``racial tensions, economic disparities, and gender-related issues'' \citep[2]{mtac_2024aug9}.} degrade ``trust in electoral processes or institutions, and [...] the target country's leadership'' \citep[6]{mtac_2024apr17}. These actors attacked both left- and right-leaning politicians \citep{mtac_2024oct23}. The state's tactics incorporated both hacking and information campaigns to maximize the effectiveness of their activities \citep[6]{mtac_2024apr17}. For example, one of the Iranian actors --- Cotton Sandstorm, an actor affiliated with the Islamic Revolutionary Guard Corps (IRGC) --- used compromised voter data to give the American public the impression that ``fraudulent ballots had been cast'' \citep[6]{mtac_2024apr17}. More recently in October, Iranian influence operations attempted to persuade voters not to cast their ballots through agents that posed as Americans against Israel \citep{mtac_2024oct23}. MTAC argued that Iran would intensity their activities shortly before the election\footnote{That is, ``within weeks or months of [the] election'' \citep[6]{mtac_2024apr17}.} because they were not as well equipped as Russia to undertake complex, continued manipulation operations \citep{mtac_2023nov8, mtac_2024oct23}. Iran's approach further differed from Russia in that it was less focused on influencing voters and engaged in efforts that were more aligned with electoral interference \citep[1]{mtac_2024aug9}.

Like Russia, there were several malicious actors affiliated with Iran that were running influence campaigns using different approaches directed at the election in November. For example, Sefid Flood's approach focused on imitating activist organizations to generate confusion, distrust in government, and degrade confidence in ``election integrity'' using ``intimidation, doxing, or violent incitement'' directed at influential individuals or organizations of a political nature \citep[2]{mtac_2024aug9}. Mint Sandstorm, an actor directed by the IRGC, targeted an important campaign official with ``a spear-phishing email'' containing a malicious link using ``a former senior advisor['s]'' hacked email credentials \citep[2]{mtac_2024aug9}. MTAC stated that these actors would likely engage in, for example, compromising ``Republican campaign targets or'' attempts to ``incite confusion, fear, or intimidation among voters in swing states'' for the 2024 election \citep[2]{mtac_2024sept17}.

China's motivations for influencing the 2024 U.S. election were shaped by the latter country's interest in maintaining peace in the Taiwan Strait, as the former intensifies attempts to establish its dominance of the region, actions which may exacerbate current tensions \citep[2]{mtac_2023nov8}. Originally, China's ``propaganda and social media disinformation'' was aimed at Taiwan, for example, during the latter's presidential elections in 2020 \citep[2]{mtac_2023nov8}. In the same year, China did not show much interest in shaping the U.S. presidential elections \citep[4]{mtac_2023nov8}. However, since extending and improving their explicit and surreptitious influence approaches (which also involve social media) over the last few years, their activities have reached countries around the world, including the U.S. during the midterms in 2022, as well as Canadian national elections \citep[2,4]{mtac_2023nov8}. Most recently, malicious China-affiliated actors have been targeting politicians critical of the state (Republicans specifically), and being supportive of their opponents in electoral competitions \citep{mtac_2024oct23}.

Like Iran, China's approaches involve exacerbating current societal cleavages and relying on ``cyber'' assets, but diverges from the former country's strategies by supporting specific news organizations and ensuring its interventions capitalize on existing political tensions to facilitate its campaigns' natural integration into public discourse \citep[4,5]{mtac_2024apr17}. When compared to Russia, while China also used AI-generated or AI-manipulated content, the latter's capacities for developing convincing ``images, memes, and videos'' surpasses the former \citep[5]{mtac_2024apr17}. 
 
Similar to Russia and Iran, there were multiple actors engaged in malicious activities that were affiliated with the Chinese government. With respect to deceptive AI-generated content, among the most productive actors is Storm-1376 (also known as Spamouflage or Taizi Flood) \citep{mtac_2024apr17,mtac_2024aug9}. One example of its activities involved leveraging a group of social media accounts numbering in the hundreds to fuel public discord surrounding ``pro-Palestinian protests at'' American post-secondary institutions \citep[7]{mtac_2024aug9}. These accounts imitated students aligned with the protests, shared pro-Palestinian content in right-leaning social media forums, re-posted geographical information about the protests from real people, and simultaneously engaged with the protests online as police and protesters collided \citep[7]{mtac_2024aug9}. Storm-1376 operated large malicious campaigns across multiple social media platforms, and was thus more delayed and less impactful in its reactions to ``political events'' than, for example, Storm-1852 \citep[7]{mtac_2024sept17}.  
 
Storm-1852 was another malicious actor affiliated with China that operated social media accounts posing as right-leaning Trump supporters and left-leaning Americans opposed to Trump \citep[6,7]{mtac_2024sept17}. This actor focused on attracting public attention online ``by re-posting content, replying to comments, polling users, and organizing `follow trains''' \citep[7]{mtac_2024sept17}.\footnote{Follow trains are a way of inorganically ``exchang[ing] follows'' on social media platforms, for example on X, actions of which are not driven by interest but rather audience expansion for a user engaging this kind of activity \citep[1017]{torres2022manufacture}. X itself does not allow this kind of activity and it constitutes as an abuse of its platform \citep[1017]{torres2022manufacture}. One of the ways that follow trains appear is through posts that almost or entirely contain mentions of other users \citep[1017]{torres2022manufacture}.} Storm-1852 relied on ``shortform video content'' that targeted both Democratic and Republican politicians and was thus most likely interested in generating apprehension and consternation in the U.S. electorate in advance of the 2024 election \citep[6,7]{mtac_2024sept17}. For example, the actor posted video content critical of President Joe Biden and Vice President Kamala Harris until their campaign was exposed, after which part of the actor's resources were no longer present on mass communication platforms \citep[6]{mtac_2024sept17}. As another example, Storm-1852 also posted videos of news coverage following the initial assassination attempt of Donald Trump, and shared available content claiming Democrats were part of the attempt \citep[7]{mtac_2024sept17}. 
 
Most of the actors mentioned in these reports have shown interest in developing and deploying AI-generated content as part of their campaigns, but a number of them reverted to using methods that have a successful track record, such as ``simple digital manipulations, mischaracterization of content, and use of trusted labels or logos atop false information'' \citep[1]{mtac_2024aug9}. MTAC details a set of features of the effects of generative AI that shape its effectiveness \citep{mtac_2024apr17}. First, ``AI-enhanced, rather than AI-generated'' video content is more persuasive and cost-effective \citep[6,7]{mtac_2024apr17}. Second, AI-generated audio is easier to develop and its lack of authenticity is more difficult to detect ``without the context clues that AI-generated video can provide'' \citep[7]{mtac_2024apr17}. Third, AI-assisted or AI-generated content is more likely to be compelling if it is disseminated through private communication channels (i.e., over the phone or private messaging) \citep[7]{mtac_2024apr17}. This deceptive content's reach on public forums, such as social media platforms, is more likely to encounter informed observers that can recognize the manipulation \citep[7]{mtac_2024apr17}. Fourth, audiences are more likely to be vulnerable to duplicitous content ``when scared or during fast-breaking events when the veracity of reported information may not yet be clear'' \citep[7]{mtac_2024apr17}. Finally, imitations of famous people are more likely to be flagged as deceptive than those of non-public facing individuals \citep[8]{mtac_2024apr17}. 

These cases highlight the importance of understanding the scale and characteristics of influence operations. Each of the malicious actors mentioned above approach manipulation campaigns in different ways, according to their interests and capacities. Understanding the motivations and capabilities of malicious actors can help intelligence communities and researchers detect these campaigns and identify the specific individuals and groups involved in their operation. 

As highlighted by this review, the most pressing concerns with respect to societal scale manipulation are the creation and spread of disinformation and networks of inauthentic personas and websites. These concerns are further exacerbated by developments in AI and its integration into the very fabric of our society. In the following section, we discuss the risks posed by the evolution and incorporation of AI into these kinds of malicious campaigns and beyond. 

\section{Evolution of AI and Societal-Scale Threats}

As AI technologies continue to evolve, their use as producers of disinformation and automated communication agents will intensify and their detection will become increasingly challenging, if not near impossible. The digital world, although in many ways an extension of our physical reality, will distort our understanding of facts and each other. Greater, rather than less, reliance on digital communication mediums will make the dilemma of the post-truth era appear insurmountable. 

In examining the broader implications of inauthentic societal-scale manipulation, the role of AI emerges as a critical concern that extends beyond current technological capabilities. While present-day manipulation largely relies on human actors wielding AI as a tool, the potential development of more advanced AI systems—particularly Artificial General Intelligence (AGI) and Artificial Superintelligence (ASI) --- presents distinct and potentially more severe risks. AGI refers to AI systems that can match or exceed human cognitive capabilities across virtually all domains, distinguishing it from today's narrow AI systems that excel only in specific tasks \citep{mclean2023risks}. ASI represents a theoretical further advancement: AI that dramatically surpasses human cognitive capabilities \citep{bostrom2020ethical, OpenAIcharter, bengio2024managing}. With such advanced capabilities, AGI/ASI could manipulate human behavior and decision-making on an unprecedented scale, pursuing their own objectives even at the expense of human welfare \citep{bengio2024managing}.

The threat posed by AGI and ASI manifests in two primary scenarios: misaligned AGI/ASI acting independently, and aligned AGI/ASI being used as a tool by human actors for harmful purposes \citep{bengio2024managing}. Here in this review, we have addressed the scenario where AI is successfully aligned, but with the values of a specific individual, group, or organization whose interests may not align with the broader welfare of citizens. In this case, the AGI/ASI can become a powerful tool for enacting the will of these actors, potentially leading to large-scale manipulation and harm. Current AI technologies, while far less advanced, already demonstrate the potential for misuse in information manipulation.\footnote{For instance, Venezuelan state media reportedly used AI-generated news anchors on a fictitious international channel to spread specific messages \citep{MITtechreview}.}

Once AGI has been achieved, it could improve the kinds of harmful content and more efficiently execute the manipulation strategies familiar to us, as well as develop new ways to shape public opinion and disrupt political and economic institutions. For example, given the fact that AI can already work faster and more efficiently at many human tasks \citep{bengio2024managing}, manipulation campaigns that involve social engineering can become even more wide-reaching, targeted, and convincing. Networks of fake personas and websites will become increasingly complex and difficult to detect. Furthermore, there may be structural, social, and psychological vulnerabilities identified by AGI that have yet to be discovered through human efforts. These vulnerabilities can also be attacked at scale with the networks of inauthentic content and personas that support the current infrastructure of manipulation campaigns. AGI can additionally be used by malicious actors to exploit structural features of social media platforms, such as recommendation algorithms, posing a much greater risk to societal cohesion than the most sophisticated automation-assisted social media accounts of today.

The potential impact of AGI/ASI on social-level manipulation represents a significant leap beyond current capabilities. While today's LLMs are already being deployed as tools in manipulation campaigns \citep{burtell2023artificial, dehnert2022persuasion}, emerging research also suggests a range of existing AI systems can exhibit deceptive and manipulative behaviors on their own \citep{park2024ai}. Moreover, \citet{salvi2024conversational} demonstrate that, in certain contexts, LLMs can already outperform human debaters, particularly when leveraging demographic insights. Progress in this domain, paired with improvements to the agency of these systems could have severe social ramifications, amplifying existing manipulation capabilities to unprecedented levels and potentially undermining democratic discourse and destabilizing social cohesion. Furthermore, efforts to incorporate agency into future systems are far from theoretical; leading AGI research organizations are actively working towards this already \citep{dailies_2024, Anthropic_2024}. The implications of such hyper-persuasive agential systems are almost boundless: from the philosophical questions it poses about what it means to be a nation-state, to more effective coercion by state actors, or even the possibility for these supercharged persuasion techniques to be leveraged to sway public opinion and policy about AI itself. Hyper-persuasive agential AI could move the needle on just about every major social issue of our time, as well as making AI governance a topic that is even harder to build popular support for.

While expert predictions about AGI development timelines vary considerably, a growing body of research suggests its potential emergence within the coming decades, underscoring the urgency of understanding and preparing for associated risks \citep{Dilmegani_2024, muller2016future, owidaitimelines}. Although the precise manifestation of AGI/ASI consequences remains uncertain, we can reason about minimum bounds of their societal impact by considering that such systems would soon possess at least human-level capabilities across all cognitive domains, while potentially exceeding them significantly \citep{barrett2017model, mclean2023risks, faroldi2024risk}. This lower bound of human-level capability, combined with advantages in speed, scale, and coordination, provides a foundation for analyzing potential manipulation risks. \citet{carroll2023characterizing} operationalized the concept of manipulation in the context of AI systems, characterizing it as dependent on factors like incentives, intent, harm, and covertness. This framework highlights the challenges in detecting and mitigating manipulation that may arise from advanced AI, even if not driven by malicious intent.

The manipulation risks posed by AGI, and more specifically, ASI systems, represent a fundamental shift from current AI-enabled threats. Rather than serving merely as tools for human actors, these advanced systems could potentially operate as autonomous agents with unprecedented capabilities for influence and control \citep{nick2014superintelligence, russell2022human}. These risks most notably manifest across the individual and social dimensions of vulnerabilities addressed in this review, each presenting unique challenges for societal resilience.

At the individual level, AGI/ASI systems could leverage capabilities for psychological manipulation that are qualitatively different from current AI systems. An analysis of superintelligent social manipulation by \citet{nick2014superintelligence} suggests that such systems would likely develop unforeseen competencies highly attuned to manipulating the psyche of the human race. This goes beyond the simple personality-based targeting or emotional manipulation seen in current systems --- an AGI/ASI would be capable of building far more sophisticated and dynamic models of human cognition than current systems. \citet[1]{rosenberg2023manipulation} analyzes adaptive influence capabilities in advanced AI systems and provides evidence for this trajectory, describing how even pre-AGI systems are likely to deploy ``real-time adaptive influence'' through virtual personas that can dynamically adjust their persuasive strategies. When scaled to AGI-level capabilities, such systems could pose unprecedented threats to human epistemic agency through their ability to deploy highly sophisticated, personalized manipulation strategies.

At a social level, the threats posed by AGI/ASI stem from fundamental challenges in AI alignment and control. The alignment risks associated with the recursive self-improvement of AI are highlighted by \citet{muehlhauser2013intelligence}, who discuss how advanced AI systems could develop sophisticated social skills to manipulate humans in service of their goals. Building on Omohundro's work on instrumental convergence --- which established how certain subgoals would likely emerge in pursuit of almost any primary objective --- we can understand advanced psychological modeling and social manipulation as likely instrumental capabilities that such systems would be motivated to develop \citep{omohundro2018basic}. \citet{sotala2018disjunctive} discusses the risks from AGI/ASI wielding advanced social manipulation capabilities, in particular, how this capability could lead to catastrophic damage even if only used on a small number of individuals if those individuals are in positions of power (e.g. heads of state). The risks of manipulative behaviors, such as deception and sycophancy, emerging during training have already been demonstrated \citep{denison2024sycophancy, park2024ai}. The potential for sophisticated social modeling and manipulation raises serious concerns about systemic vulnerabilities --— particularly when combined with other vectors like cyber or biological capabilities --— as manipulation of even key individuals could enable catastrophic outcomes.
These theoretical frameworks suggest that AGI/ASI manipulation capabilities would represent a fundamental departure from current techniques, characterized by total integration of psychological modeling, prediction, and intervention at speeds and scales beyond human comprehension. 

Although the state of technological advancements in AI have not yet reached the point of no return, we must develop safeguards to ensure that its unfettered use is not possible. In the conclusions below, we integrate the observations noted in this review and discuss potential mitigation strategies for the threats we have addressed. 

\section{Conclusions}

In our review, we distinguished between misinformation and disinformation on the basis of harmful intent. However, intent is not as pertinent to the mitigation of societally harmful content as the inaccuracy of the information shared. In order to prevent the spread of harmful content, developing tools for verifying information as it spreads online in real time is paramount. In line with this objective, one of the core objectives of our research team, the Complex Data Lab, has been misinformation detection \citep{tian2024web, vergho2024comparing, pelrine2023towards}. In our most recent project, we combine the efforts of an LLM agent with a web search agent to detect misinformation with greater accuracy \citep{tian2024web}. 

The distinction between misinformation and disinformation is more semantic than substantive from the perspective of policymakers. Intent does matter from disciplinary perspective, as regular people that do not consciously participate in manipulation campaigns should not be subject to the repercussions that malicious actors may face. Producers and deliberate spreaders\footnote{In other words, if these individuals were aware that they were sharing inaccurate information, harmful intent is likely present.} of disinformation should be the central concerns of emerging accountability policies. 

From a policymaking perspective, with respect to mitigating the proliferation of harmful content, all of the different types of content described in this review are forms of disinformation or malinformation. The central difference between these two concepts relates to the authenticity of the content; while disinformation is false, malinformation is based in truth but presented in a manipulative or distorted way. Fake news and deepfakes fall into the category of disinformation because they involve fabricated pieces of information. Conspiracy theories, propaganda, rumors, and toxic statements could be either disinformation or malinformation, depending on whether or not they are substantiated. Toxic behavior is unique in that while it involves subjective commentary and negativity, it can also contain claims that may or may not encompass elements of truth or fiction. It is also possible that each of the types of harmful content we address can be partially truthful, in which case the difference between disinformation and malinformation almost becomes a moot point. 

Understanding the various forms of harmful content is thus more relevant for developing policies aimed to hold people accountable for the information they create and share. Each of these different forms require varying levels of resources, which not only reflects the capacities of malicious actors, but also their conviction. For example, setting up a fake news website requires more effort and intentional deception than intentionally spreading rumors or engaging in toxic speech. Methods of deployment that involve networks of fake assets also require greater investment and coordination, and pose societal-level risks, relative to individual malicious actors engaging in manipulation efforts on their own.

In order to spread or effectively use a piece of harmful content, we have discussed how individual or networks of inauthentic personas and websites are integral to achieve societal-level disruption. These two assets can be used to both manipulate public opinion and obtain sensitive information. To manipulate public opinion on social media, inauthentic personas, such as automated, partially automated or human-operated accounts, help facilitate the spread of harmful content. 

With respect to the vulnerabilities we have identified, policymakers should address each of the structural, social, and psychological levels in order to develop effective countermeasures. For example, social media companies could restrain the spread of unverified information by temporarily reducing the visibility of users of their Application Program Interfaces, as these users are at least partially automated. Given that automated accounts help facilitate the spread of information, this measure would likely be effective. As another example, a potential countermeasure to social and psychological vulnerabilities is education. Equipping people with the knowledge of how to verify and moderate their reactions to novel or negative information would help reduce the rate at which harmful content spreads.

However, any countermeasures would fall short if policymakers do not consider the use and progression of AI. As discussed through the case studies, although the technological state of AI has not yet reached the point where it is integral to manipulation campaigns, governments' current and projected mitigation capacities fall short of the present state and pace of AI evolution \citep{bengio2024managing}. These technologies will continue to reinforce the effectiveness of these manipulation strategies, and without mitigation efforts that match the potency and speed of their development, we will truly be living in a post-truth era. 

From content to deployment, manipulation strategies by malicious actors threaten the stability of societies. Social media has become an integral testing ground for ``modern warfare'' and constitutes as a major threat to cybersecurity for countries around the world \citep[1]{chen2022social}. The evolution of AI also poses severe risks to content generation and dissemination for which we have yet to implement safeguards against. These risks may be approaching at a pace that we may not be able to prepare for without fundamental changes to public policy, regulation of technology companies, and international collaboration. One of the founders of AI, Yoshua Bengio, has predicted that in the worst case scenario, we may achieve AGI ``within the next few years,'' if not, within twenty years \citep[1]{bengio2024government}. According to \citet[59]{csis2023}, AI will likely enhance existing attack vectors, specifically ``disinformation, the targeting of government/military personnel by adversarial forces, phishing/social engineering, and mimicking biometric data.'' As these strategies improve over time, so too must our approaches to combating them. 

Intelligence agencies and researchers have identified several mitigation strategies. For example, the Canadian Security Intelligence Service (CSIS) encourages the development of AI-generated audiovisual and text detection capabilities \citep{csis2023}. Although the detection of AI content is becoming increasingly challenging, keeping pace with developments will make us more likely to be able to grapple with the problem.

Another potentially promising direction for research related to the use of blockchain technologies for content verification \citep{csis2023}. An immutable record of all changes to original audiovisual and text content from its point of origin will not only track chain-of-custody, but could also inspire confidence among professional and public circles in their use of this content in the context of, for example, legal proceedings \citep{csis2023}. 

An additional approach that is more proactive involves the development of, for example, public policies that require explicit attribution of AI's use in content development, as well as legislation that bears ramifications for actors using this content in malicious ways \citep{csis2023}. Europe has been at the forefront of legislative developments with respect to AI content, recently putting forth the European Artificial Intelligence Act \citep{EUAIAct2024}. 

Beyond the responsibilities of domestic public agencies and institutions, ``international collaboration'' that brings countries around the world to the table to discuss pertinent issues relating to AI, such as blockchain-based verification, will be integral given the borderless nature of digital space \citep[53]{csis2023}. Nationally and internationally, governments should develop policies and practices that allow them to respond to AI developments quickly and effectively \citep{bengio2024managing}. They should be able to incorporate expert feedback and adapt to changes in the technological landscape as they occur \citep{bengio2024managing}. Although international competition may serve as an obstacle to collaboration, agreements can be established to overcome it \citep{bengio2024managing}. The focus of authorities should be on actors developing the most advanced AI models so as not to stifle innovation \citep{bengio2024managing} through, for example, open-source projects. Furthermore, governments should have a certain level of oversight with respect to the development of these models to be able to foresee potential challenges \citep{bengio2024managing}. 

Since corrections to misconceptions held by members of the general public are most effective when executed by ``someone they know'' or ``influencers'' \citep[27,53]{csis2023}, there is an important socialization component to information sharing.\footnote{According to \citet[394]{ross2019social}, ``the most central actor in [a] network [often] determines the final consensus''.}  Educational programs that inform users and inspire public discussion online about the pitfalls of not verifying information will likely be effective \citep{csis2023, fazio2024combating}. One of the educational approaches that have proven to be successful involves the concept of ``inoculation'' \citep{mannix2022personal, fazio2024combating}. Inoculation entails ``exposing'' the public to the kinds of content and strategies they may encounter during real manipulation campaigns ``to bolster their psychological resistance'' \citep[333]{mannix2022personal}. It has been found that exposure to strategies --- for example, ``emotionally manipulative language, polarizing language, conspiratorial reasoning, trolling, and logical fallacies'' --- is more effective than exposing individuals to harmful content \citep[333]{mannix2022personal}. 

Finally, technology companies, such as OpenAI, X, or Facebook should be held responsible for the content shared on their platforms \citep{csis2023}. This entails that content and users be reported and restricted from using these platforms in a timely manner \citep{csis2023}. This recommendation should be implemented with great care, especially in democratic countries that have a duty to their citizens to protect open communication. 

This review paper was motivated by the evolving threat landscape of digital space. Proactive and reactive mitigation approaches will be imperative as we prepare for the emerging threats on the horizon, with harmful content and malicious actors powered by AGI. 

\newpage

\bibliographystyle{plainnat}
\bibliography{bibliography}

\begin{thebibliography}{148}
\providecommand{\natexlab}[1]{#1}
\providecommand{\url}[1]{\texttt{#1}}
\expandafter\ifx\csname urlstyle\endcsname\relax
  \providecommand{\doi}[1]{doi: #1}\else
  \providecommand{\doi}{doi: \begingroup \urlstyle{rm}\Url}\fi

\bibitem[A{\"\i}meur et~al.(2023)A{\"\i}meur, Amri, and Brassard]{aimeur2023fake}
Esma A{\"\i}meur, Sabrine Amri, and Gilles Brassard.
\newblock Fake news, disinformation and misinformation in social media: a review.
\newblock \emph{Social Network Analysis and Mining}, 13\penalty0 (1):\penalty0 30, 2023.

\bibitem[Akhtar et~al.(2024)Akhtar, Masood, Ikram, and Kanhere]{akhtar2024sok}
Mohammad~Majid Akhtar, Rahat Masood, Muhammad Ikram, and Salil~S Kanhere.
\newblock Sok: False information, bots and malicious campaigns: Demystifying elements of social media manipulations.
\newblock In \emph{Proceedings of the 19th ACM Asia Conference on Computer and Communications Security}, pages 1784--1800, 2024.

\bibitem[Alabdan(2020)]{alabdan2020phishing}
Rana Alabdan.
\newblock Phishing attacks survey: Types, vectors, and technical approaches.
\newblock \emph{Future internet}, 12\penalty0 (10):\penalty0 168, 2020.

\bibitem[Ali and Mohd~Zaharon(2024)]{ali2024phishing}
Mazurina~Mohd Ali and Nur~Farhana Mohd~Zaharon.
\newblock Phishing—a cyber fraud: The types, implications and governance.
\newblock \emph{International Journal of Educational Reform}, 33\penalty0 (1):\penalty0 101--121, 2024.

\bibitem[Alkhalil et~al.(2021)Alkhalil, Hewage, Nawaf, and Khan]{alkhalil2021phishing}
Zainab Alkhalil, Chaminda Hewage, Liqaa Nawaf, and Imtiaz Khan.
\newblock Phishing attacks: A recent comprehensive study and a new anatomy.
\newblock \emph{Frontiers in Computer Science}, 3:\penalty0 563060, 2021.

\bibitem[Alonso-Villota and Arcos(2024)]{alonso2024coercion}
Marina Alonso-Villota and Rub{\'e}n Arcos.
\newblock The coercion-manipulation-persuasion framework: Analyzing the modus operandi of systems of non-state actors.
\newblock \emph{Terrorism and Political Violence}, pages 1--19, 2024.

\bibitem[Anthropic(2024)]{Anthropic_2024}
Anthropic.
\newblock Introducing computer use, a new claude 3.5 sonnet, and claude 3.5 haiku, 10 2024.
\newblock URL \url{https://www.anthropic.com/news/3-5-models-and-computer-use}.

\bibitem[Avalle et~al.(2024)Avalle, Di~Marco, Etta, Sangiorgio, Alipour, Bonetti, Alvisi, Scala, Baronchelli, Cinelli, et~al.]{avalle2024persistent}
Michele Avalle, Niccol{\`o} Di~Marco, Gabriele Etta, Emanuele Sangiorgio, Shayan Alipour, Anita Bonetti, Lorenzo Alvisi, Antonio Scala, Andrea Baronchelli, Matteo Cinelli, et~al.
\newblock Persistent interaction patterns across social media platforms and over time.
\newblock \emph{Nature}, 628\penalty0 (8008):\penalty0 582--589, 2024.

\bibitem[Barrett and Baum(2017)]{barrett2017model}
Anthony~M Barrett and Seth~D Baum.
\newblock A model of pathways to artificial superintelligence catastrophe for risk and decision analysis.
\newblock \emph{Journal of Experimental \& Theoretical Artificial Intelligence}, 29\penalty0 (2):\penalty0 397--414, 2017.

\bibitem[Bayer and Rupert(2004)]{bayer2004effects}
Resat Bayer and Matthew~C Rupert.
\newblock Effects of civil wars on international trade, 1950-92.
\newblock \emph{Journal of Peace Research}, 41\penalty0 (6):\penalty0 699--713, 2004.

\bibitem[Bengio(2024)]{bengio2024government}
Yoshua Bengio.
\newblock Government interventions to avert future catastrophic ai risks.
\newblock \emph{Harvard Data Science Review}, \penalty0 (Special Issue 5), 2024.

\bibitem[Bengio et~al.(2024)Bengio, Hinton, Yao, Song, Abbeel, Darrell, Harari, Zhang, Xue, Shalev-Shwartz, et~al.]{bengio2024managing}
Yoshua Bengio, Geoffrey Hinton, Andrew Yao, Dawn Song, Pieter Abbeel, Trevor Darrell, Yuval~Noah Harari, Ya-Qin Zhang, Lan Xue, Shai Shalev-Shwartz, et~al.
\newblock Managing extreme ai risks amid rapid progress.
\newblock \emph{Science}, 384\penalty0 (6698):\penalty0 842--845, 2024.
\newblock URL \url{https://www.science.org/doi/full/10.1126/science.adn0117}.

\bibitem[Berinsky(2017)]{berinsky2017rumors}
Adam~J Berinsky.
\newblock Rumors and health care reform: Experiments in political misinformation.
\newblock \emph{British journal of political science}, 47\penalty0 (2):\penalty0 241--262, 2017.

\bibitem[Bhardwaj et~al.(2020)Bhardwaj, Sapra, Kumar, Kumar, and Arthi]{bhardwaj2020phishing}
Akashdeep Bhardwaj, Varun Sapra, Aman Kumar, Naman Kumar, and S~Arthi.
\newblock Why is phishing still successful?
\newblock \emph{Computer Fraud \& Security}, 2020\penalty0 (9):\penalty0 15--19, 2020.

\bibitem[Blais(2024)]{dejardin2024}
Stéphane Blais.
\newblock Quebec police arrest three in \$9-million fraud, data theft case involving desjardins, 06 2024.
\newblock URL \url{https://globalnews.ca/news/10563079/desjardins-data-theft-arrests/}.

\bibitem[Bossetta(2018)]{bossetta2018weaponization}
Michael Bossetta.
\newblock The weaponization of social media: Spear phishing and cyberattacks on democracy.
\newblock \emph{Journal of international affairs}, 71\penalty0 (1.5):\penalty0 97--106, 2018.

\bibitem[Bostrom(2014)]{nick2014superintelligence}
Nick Bostrom.
\newblock \emph{Superintelligence: Paths, dangers, strategies}.
\newblock Oxford University Press, Oxford, 2014.

\bibitem[Bostrom(2020)]{bostrom2020ethical}
Nick Bostrom.
\newblock Ethical issues in advanced artificial intelligence.
\newblock \emph{Machine Ethics and Robot Ethics}, pages 69--75, 2020.

\bibitem[Bradshaw et~al.(2021)Bradshaw, Bailey, and Howard]{bradshaw2021industrialized}
Samantha Bradshaw, Hannah Bailey, and Philip~N Howard.
\newblock \emph{Industrialized disinformation: 2020 global inventory of organized social media manipulation}.
\newblock Computational Propaganda Project at the Oxford Internet Institute, 2021.
\newblock URL \url{https://demtech.oii.ox.ac.uk/wp-content/uploads/sites/12/2021/02/CyberTroop-Report20-Draft9.pdf}.

\bibitem[Broda and Str{\"o}mb{\"a}ck(2024)]{broda2024misinformation}
Elena Broda and Jesper Str{\"o}mb{\"a}ck.
\newblock Misinformation, disinformation, and fake news: lessons from an interdisciplinary, systematic literature review.
\newblock \emph{Annals of the International Communication Association}, 48\penalty0 (2):\penalty0 139--166, 2024.

\bibitem[Burtell and Woodside(2023)]{burtell2023artificial}
Matthew Burtell and Thomas Woodside.
\newblock Artificial influence: An analysis of ai-driven persuasion.
\newblock \emph{arXiv preprint arXiv:2303.08721}, 2023.

\bibitem[{Cambridge Dictionary}(2024)]{malicious}
{Cambridge Dictionary}.
\newblock malicious, 2024.
\newblock URL \url{https://dictionary.cambridge.org/dictionary/english/malicious}.

\bibitem[Carrasco-Farr{\'e}(2022)]{carrasco2022fingerprints}
Carlos Carrasco-Farr{\'e}.
\newblock The fingerprints of misinformation: how deceptive content differs from reliable sources in terms of cognitive effort and appeal to emotions.
\newblock \emph{Humanities and Social Sciences Communications}, 9\penalty0 (1):\penalty0 1--18, 2022.

\bibitem[Carroll et~al.(2023)Carroll, Chan, Ashton, and Krueger]{carroll2023characterizing}
Micah Carroll, Alan Chan, Henry Ashton, and David Krueger.
\newblock Characterizing manipulation from ai systems.
\newblock In \emph{Proceedings of the 3rd ACM Conference on Equity and Access in Algorithms, Mechanisms, and Optimization}, pages 1--13, 2023.

\bibitem[Chang et~al.(2023)Chang, Golightly, Xu, Boonmee, and Liu]{chang2023cybersecurity}
Victor Chang, Lewis Golightly, Qianwen~Ariel Xu, Thanaporn Boonmee, and Ben~S Liu.
\newblock Cybersecurity for children: an investigation into the application of social media.
\newblock \emph{Enterprise Information Systems}, 17\penalty0 (11):\penalty0 2188122, 2023.

\bibitem[Chen et~al.(2022)Chen, Chen, and Xia]{chen2022social}
Long Chen, Jianguo Chen, and Chunhe Xia.
\newblock Social network behavior and public opinion manipulation.
\newblock \emph{Journal of Information Security and Applications}, 64:\penalty0 103060, 2022.

\bibitem[Cheng et~al.(2017)Cheng, Bernstein, Danescu-Niculescu-Mizil, and Leskovec]{cheng2017anyone}
Justin Cheng, Michael Bernstein, Cristian Danescu-Niculescu-Mizil, and Jure Leskovec.
\newblock Anyone can become a troll: Causes of trolling behavior in online discussions.
\newblock In \emph{Proceedings of the 2017 ACM conference on computer supported cooperative work and social computing}, pages 1217--1230, 2017.

\bibitem[Chow and Kono(2017)]{chow2017entry}
Wilfred~Ming Chow and Daniel~Yuichi Kono.
\newblock Entry, vulnerability, and trade policy: Why some autocrats like international trade.
\newblock \emph{International Studies Quarterly}, 61\penalty0 (4):\penalty0 892--906, 2017.

\bibitem[CISA(2022)]{CISA_2022}
CISA.
\newblock Tactics of disinformation, 10 2022.
\newblock URL \url{https://www.cisa.gov/sites/default/files/publications/tactics-of-disinformation_508.pdf}.

\bibitem[Confessore(2018)]{cambridgelondon2018}
Nicholas Confessore.
\newblock Cambridge analytica and facebook: The scandal and the fallout so far, 04 2018.
\newblock URL \url{https://www.nytimes.com/2018/04/04/us/politics/cambridge-analytica-scandal-fallout.html}.

\bibitem[Cook et~al.(2014)Cook, Waugh, Abdipanah, Hashemi, and Rahman]{cook2014twitter}
David~M Cook, Benjamin Waugh, Maldini Abdipanah, Omid Hashemi, and Shaquille~Abdul Rahman.
\newblock Twitter deception and influence: Issues of identity, slacktivism, and puppetry.
\newblock \emph{Journal of Information Warfare}, 13\penalty0 (1):\penalty0 58--71, 2014.

\bibitem[Cooke et~al.(2024)Cooke, Edwards, Barkoff, and Kelly]{cooke2024good}
Di~Cooke, Abigail Edwards, Sophia Barkoff, and Kathryn Kelly.
\newblock As good as a coin toss human detection of ai-generated images, videos, audio, and audiovisual stimuli.
\newblock \emph{arXiv preprint arXiv:2403.16760}, 2024.

\bibitem[Cresci(2020)]{cresci2020decade}
Stefano Cresci.
\newblock A decade of social bot detection.
\newblock \emph{Communications of the ACM}, 63\penalty0 (10):\penalty0 72--83, 2020.

\bibitem[CSIS(2023)]{csis2023}
CSIS.
\newblock The evolution of disinformation: A deepfake future, 10 2023.
\newblock URL \url{https://www.canada.ca/content/dam/csis-scrs/documents/publications/2023/The\%20Evolution\%20of\%20Disinformation\%20-\%20Deepfake\%20Report_EN_DIGITAL.pdf}.

\bibitem[De~Angelis et~al.(2023)De~Angelis, Baglivo, Arzilli, Privitera, Ferragina, Tozzi, and Rizzo]{de2023chatgpt}
Luigi De~Angelis, Francesco Baglivo, Guglielmo Arzilli, Gaetano~Pierpaolo Privitera, Paolo Ferragina, Alberto~Eugenio Tozzi, and Caterina Rizzo.
\newblock Chatgpt and the rise of large language models: the new ai-driven infodemic threat in public health.
\newblock \emph{Frontiers in public health}, 11:\penalty0 1166120, 2023.

\bibitem[De~Salve et~al.(2019)De~Salve, Mori, Guidi, and Ricci]{de2019analysis}
Andrea De~Salve, Paolo Mori, Barbara Guidi, and Laura Ricci.
\newblock An analysis of the internal organization of facebook groups.
\newblock \emph{IEEE Transactions on Computational Social Systems}, 6\penalty0 (6):\penalty0 1245--1256, 2019.

\bibitem[Dehnert and Mongeau(2022)]{dehnert2022persuasion}
Marco Dehnert and Paul~A Mongeau.
\newblock Persuasion in the age of artificial intelligence (ai): Theories and complications of ai-based persuasion.
\newblock \emph{Human Communication Research}, 48\penalty0 (3):\penalty0 386--403, 2022.

\bibitem[Del~Vicario et~al.(2016)Del~Vicario, Bessi, Zollo, Petroni, Scala, Caldarelli, Stanley, and Quattrociocchi]{del2016spreading}
Michela Del~Vicario, Alessandro Bessi, Fabiana Zollo, Fabio Petroni, Antonio Scala, Guido Caldarelli, H~Eugene Stanley, and Walter Quattrociocchi.
\newblock The spreading of misinformation online.
\newblock \emph{Proceedings of the national academy of Sciences}, 113\penalty0 (3):\penalty0 554--559, 2016.

\bibitem[Del~Vicario et~al.(2017)Del~Vicario, Scala, Caldarelli, Stanley, and Quattrociocchi]{del2017modeling}
Michela Del~Vicario, Antonio Scala, Guido Caldarelli, H~Eugene Stanley, and Walter Quattrociocchi.
\newblock Modeling confirmation bias and polarization.
\newblock \emph{Scientific reports}, 7\penalty0 (1):\penalty0 40391, 2017.

\bibitem[Denison et~al.(2024)Denison, MacDiarmid, Barez, Duvenaud, Kravec, Marks, Schiefer, Soklaski, Tamkin, Kaplan, et~al.]{denison2024sycophancy}
Carson Denison, Monte MacDiarmid, Fazl Barez, David Duvenaud, Shauna Kravec, Samuel Marks, Nicholas Schiefer, Ryan Soklaski, Alex Tamkin, Jared Kaplan, et~al.
\newblock Sycophancy to subterfuge: Investigating reward-tampering in large language models.
\newblock \emph{arXiv preprint arXiv:2406.10162}, 2024.

\bibitem[Dilmegani(2024)]{Dilmegani_2024}
Cem Dilmegani.
\newblock When will singularity happen? 1700 expert opinions of agi, 11 2024.
\newblock URL \url{https://research.aimultiple.com/artificial-general-intelligence-singularity-timing/}.

\bibitem[Douglas(2016)]{douglas2016doxing}
David~M Douglas.
\newblock Doxing: a conceptual analysis.
\newblock \emph{Ethics and information technology}, 18\penalty0 (3):\penalty0 199--210, 2016.

\bibitem[Ecker et~al.(2022)Ecker, Lewandowsky, Cook, Schmid, Fazio, Brashier, Kendeou, Vraga, and Amazeen]{ecker2022psychological}
Ullrich~KH Ecker, Stephan Lewandowsky, John Cook, Philipp Schmid, Lisa~K Fazio, Nadia Brashier, Panayiota Kendeou, Emily~K Vraga, and Michelle~A Amazeen.
\newblock The psychological drivers of misinformation belief and its resistance to correction.
\newblock \emph{Nature Reviews Psychology}, 1\penalty0 (1):\penalty0 13--29, 2022.

\bibitem[EU(2024)]{EUAIAct2024}
EU.
\newblock Regulation - eu - 2024/1689 - en - eur-lex, 06 2024.
\newblock URL \url{http://data.europa.eu/eli/reg/2024/1689/oj}.

\bibitem[Eysenbach et~al.(2020)]{eysenbach2020fight}
Gunther Eysenbach et~al.
\newblock How to fight an infodemic: the four pillars of infodemic management.
\newblock \emph{Journal of medical Internet research}, 22\penalty0 (6):\penalty0 e21820, 2020.

\bibitem[Falowo et~al.(2022)Falowo, Popoola, Riep, Adewopo, and Koch]{falowo2022threat}
Olufunsho~I Falowo, Saheed Popoola, Josette Riep, Victor~A Adewopo, and Jacob Koch.
\newblock Threat actors’ tenacity to disrupt: Examination of major cybersecurity incidents.
\newblock \emph{IEEE Access}, 10:\penalty0 134038--134051, 2022.

\bibitem[Faroldi(2024)]{faroldi2024risk}
Federico~LG Faroldi.
\newblock Risk and artificial general intelligence.
\newblock \emph{AI \& SOCIETY}, pages 1--9, 2024.

\bibitem[Fazio et~al.(2024)Fazio, Rand, Lewandowsky, Susmann, Berinsky, Guess, Kendeou, Lyons, Miller, Newman, and et~al.]{fazio2024combating}
Lisa Fazio, David~G Rand, Stephan Lewandowsky, Mark Susmann, Adam~J Berinsky, Andrew~M Guess, Panayiota Kendeou, Benjamin Lyons, Joanne~M Miller, Eryn Newman, and et~al.
\newblock Combating misinformation: A megastudy of nine interventions designed to reduce the sharing of and belief in false and misleading headlines.
\newblock \emph{PsyArXiv preprint PsyArXiv:10.31234}, 2024.
\newblock URL \url{https://osf.io/preprints/psyarxiv/uyjha}.

\bibitem[FBI(2020)]{ransomware2020}
FBI.
\newblock Ransomware, 04 2020.
\newblock URL \url{https://www.fbi.gov/how-we-can-help-you/scams-and-safety/common-frauds-and-scams/ransomware}.

\bibitem[Ferrara(2023)]{ferrara2023social}
Emilio Ferrara.
\newblock Social bot detection in the age of chatgpt: Challenges and opportunities.
\newblock \emph{First Monday}, 2023.

\bibitem[Frank et~al.(2024)Frank, Herbert, Ricker, Sch{\"o}nherr, Eisenhofer, Fischer, D{\"u}rmuth, and Holz]{frank2024representative}
Joel Frank, Franziska Herbert, Jonas Ricker, Lea Sch{\"o}nherr, Thorsten Eisenhofer, Asja Fischer, Markus D{\"u}rmuth, and Thorsten Holz.
\newblock A representative study on human detection of artificially generated media across countries.
\newblock In \emph{2024 IEEE Symposium on Security and Privacy (SP)}, pages 55--73. IEEE, 2024.

\bibitem[Gallotti et~al.(2020)Gallotti, Valle, Castaldo, Sacco, and De~Domenico]{gallotti2020assessing}
Riccardo Gallotti, Francesco Valle, Nicola Castaldo, Pierluigi Sacco, and Manlio De~Domenico.
\newblock Assessing the risks of ‘infodemics’ in response to covid-19 epidemics.
\newblock \emph{Nature human behaviour}, 4\penalty0 (12):\penalty0 1285--1293, 2020.

\bibitem[Giglietto et~al.(2020)Giglietto, Righetti, Rossi, and Marino]{giglietto2020takes}
Fabio Giglietto, Nicola Righetti, Luca Rossi, and Giada Marino.
\newblock It takes a village to manipulate the media: coordinated link sharing behavior during 2018 and 2019 italian elections.
\newblock \emph{Information, Communication \& Society}, 23\penalty0 (6):\penalty0 867--891, 2020.

\bibitem[Goldman and Baker(2019)]{goldman2019free}
Alvin~I Goldman and Daniel Baker.
\newblock Free speech, fake news, and democracy.
\newblock \emph{First Amend. L. Rev.}, 18:\penalty0 66, 2019.

\bibitem[Guess and Lyons(2020)]{guess2020misinformation}
Andrew~M Guess and Benjamin~A Lyons.
\newblock Misinformation, disinformation, and online propaganda.
\newblock \emph{Social media and democracy: The state of the field, prospects for reform}, 10, 2020.

\bibitem[Hermann(2023)]{hermann2023psychological}
Erik Hermann.
\newblock Psychological targeting: nudge or boost to foster mindful and sustainable consumption?
\newblock \emph{AI \& SOCIETY}, 38\penalty0 (2):\penalty0 961--962, 2023.

\bibitem[Himelein-Wachowiak et~al.(2021)Himelein-Wachowiak, Giorgi, Devoto, Rahman, Ungar, Schwartz, Epstein, Leggio, and Curtis]{himelein2021bots}
McKenzie Himelein-Wachowiak, Salvatore Giorgi, Amanda Devoto, Muhammad Rahman, Lyle Ungar, H~Andrew Schwartz, David~H Epstein, Lorenzo Leggio, and Brenda Curtis.
\newblock Bots and misinformation spread on social media: Implications for covid-19.
\newblock \emph{Journal of medical Internet research}, 23\penalty0 (5):\penalty0 e26933, 2021.

\bibitem[Hodgson(2006)]{hodgson2006institutions}
Geoffrey~M Hodgson.
\newblock What are institutions?
\newblock \emph{Journal of economic issues}, 40\penalty0 (1):\penalty0 1--25, 2006.

\bibitem[{IC3}(2024)]{usic_insiderthreats}
{IC3}.
\newblock 2024 u.s. federal elections: The insider threat, 06 2024.
\newblock URL \url{https://www.ic3.gov/Media/News/2024/240628.pdf}.

\bibitem[Innes et~al.(2021)Innes, Dobreva, and Innes]{innes2021disinformation}
Martin Innes, Diyana Dobreva, and Helen Innes.
\newblock Disinformation and digital influencing after terrorism: Spoofing, truthing and social proofing.
\newblock \emph{Contemporary Social Science}, 16\penalty0 (2):\penalty0 241--255, 2021.

\bibitem[Isaak and Hanna(2018)]{isaak2018user}
Jim Isaak and Mina~J Hanna.
\newblock User data privacy: Facebook, cambridge analytica, and privacy protection.
\newblock \emph{Computer}, 51\penalty0 (8):\penalty0 56--59, 2018.

\bibitem[Johnson and Fowler(2011)]{johnson2011evolution}
Dominic~DP Johnson and James~H Fowler.
\newblock The evolution of overconfidence.
\newblock \emph{Nature}, 477\penalty0 (7364):\penalty0 317--320, 2011.

\bibitem[Kandel(2020)]{kandel2020information}
Nirmal Kandel.
\newblock Information disorder syndrome and its management.
\newblock \emph{JNMA: Journal of the Nepal Medical Association}, 58\penalty0 (224):\penalty0 280, 2020.

\bibitem[Keller and Klinger(2019)]{keller2019social}
Tobias~R Keller and Ulrike Klinger.
\newblock Social bots in election campaigns: Theoretical, empirical, and methodological implications.
\newblock \emph{Political Communication}, 36\penalty0 (1):\penalty0 171--189, 2019.

\bibitem[Khaund et~al.(2021)Khaund, Kirdemir, Agarwal, Liu, and Morstatter]{khaund2021social}
Tuja Khaund, Baris Kirdemir, Nitin Agarwal, Huan Liu, and Fred Morstatter.
\newblock Social bots and their coordination during online campaigns: a survey.
\newblock \emph{IEEE Transactions on Computational Social Systems}, 9\penalty0 (2):\penalty0 530--545, 2021.

\bibitem[Kietzmann et~al.(2020)Kietzmann, Lee, McCarthy, and Kietzmann]{kietzmann2020deepfakes}
Jan Kietzmann, Linda~W Lee, Ian~P McCarthy, and Tim~C Kietzmann.
\newblock Deepfakes: Trick or treat?
\newblock \emph{Business Horizons}, 63\penalty0 (2):\penalty0 135--146, 2020.

\bibitem[Krafft and Donovan(2020)]{krafft2020disinformation}
Peaks~M Krafft and Joan Donovan.
\newblock Disinformation by design: The use of evidence collages and platform filtering in a media manipulation campaign.
\newblock \emph{Political Communication}, 37\penalty0 (2):\penalty0 194--214, 2020.

\bibitem[Krotoszynski~Jr(2003)]{krotoszynski2003comparative}
Ronald~J Krotoszynski~Jr.
\newblock A comparative perspective on the first amendment: free speech, militant democracy, and the primacy of dignity as a preferred constitutional value in germany.
\newblock \emph{Tul. L. Rev.}, 78:\penalty0 1549, 2003.

\bibitem[Kuklinski et~al.(2000)Kuklinski, Quirk, Jerit, Schwieder, and Rich]{kuklinski2000misinformation}
James~H Kuklinski, Paul~J Quirk, Jennifer Jerit, David Schwieder, and Robert~F Rich.
\newblock Misinformation and the currency of democratic citizenship.
\newblock \emph{The Journal of Politics}, 62\penalty0 (3):\penalty0 790--816, 2000.

\bibitem[La et~al.(2022)La, Tran, Tran, Tran, Dang-Nguyen, and Dao]{la2022multimodal}
Tuan-Vinh La, Quang-Tien Tran, Thanh-Phuc Tran, Anh-Duy Tran, Duc-Tien Dang-Nguyen, and Minh-Son Dao.
\newblock Multimodal cheapfakes detection by utilizing image captioning for global context.
\newblock In \emph{Proceedings of the 3rd ACM Workshop on Intelligent Cross-Data Analysis and Retrieval}, pages 9--16, 2022.

\bibitem[Latkin et~al.(2023)Latkin, Dayton, Strickland, Colon, Rimal, and Boodram]{latkin2023assessment}
Carl~A Latkin, Lauren Dayton, Justin~C Strickland, Brian Colon, Rajiv Rimal, and Basmattee Boodram.
\newblock An assessment of the rapid decline of trust in us sources of public information about covid-19.
\newblock In \emph{Vaccine Communication in a Pandemic}, pages 22--31. Routledge, 2023.

\bibitem[Lazer et~al.(2018)Lazer, Baum, Benkler, Berinsky, Greenhill, Menczer, Metzger, Nyhan, Pennycook, Rothschild, et~al.]{lazer2018science}
David~MJ Lazer, Matthew~A Baum, Yochai Benkler, Adam~J Berinsky, Kelly~M Greenhill, Filippo Menczer, Miriam~J Metzger, Brendan Nyhan, Gordon Pennycook, David Rothschild, et~al.
\newblock The science of fake news.
\newblock \emph{Science}, 359\penalty0 (6380):\penalty0 1094--1096, 2018.

\bibitem[Lee et~al.(2014)Lee, Yen, and Hsiao]{lee2014understanding}
Maria~R Lee, David~C Yen, and CY~Hsiao.
\newblock Understanding the perceived community value of facebook users.
\newblock \emph{Computers in Human Behavior}, 35:\penalty0 350--358, 2014.

\bibitem[Lee(2023)]{taiwanintelligence}
Yimou Lee.
\newblock Taiwan intelligence says china leadership discussed election interference, 12 2023.
\newblock URL \url{https://www.reuters.com/world/asia-pacific/taiwan-intelligence-says-china-leadership-meet-election-interference-2023-12-08/}.

\bibitem[Mannix(2022)]{mannix2022personal}
Lauren Mannix.
\newblock Personal data exploitation and social media manipulation as a security threat for nato nations and democratic societies.
\newblock \emph{Journal of Military and Strategic Studies}, 22\penalty0 (1), 2022.

\bibitem[Mansfield et~al.(2017)Mansfield, Milner, and Rosendorff]{mansfield2017democracies}
Edward~D Mansfield, Helen~V Milner, and B~Peter Rosendorff.
\newblock Why democracies cooperate more: Electoral control and international trade agreements.
\newblock In \emph{Global Trade}, pages 215--252. Routledge, 2017.

\bibitem[Marengo et~al.(2023)Marengo, Elhai, and Montag]{marengo2023predicting}
Davide Marengo, Jon~D Elhai, and Christian Montag.
\newblock Predicting big five personality traits from smartphone data: a meta-analysis on the potential of digital phenotyping.
\newblock \emph{Journal of Personality}, 91\penalty0 (6):\penalty0 1410--1424, 2023.

\bibitem[Mathur et~al.(2023)Mathur, Wang, Schwemmer, Hamin, Stewart, and Narayanan]{mathur2023manipulative}
Arunesh Mathur, Angelina Wang, Carsten Schwemmer, Maia Hamin, Brandon~M Stewart, and Arvind Narayanan.
\newblock Manipulative tactics are the norm in political emails: Evidence from 300k emails from the 2020 us election cycle.
\newblock \emph{Big Data \& Society}, 10\penalty0 (1):\penalty0 20539517221145371, 2023.

\bibitem[McHatton and Ghazinour(2023)]{mchatton2023mitigating}
Jeremy McHatton and Kambiz Ghazinour.
\newblock Mitigating social media privacy concerns-a comprehensive study.
\newblock In \emph{Proceedings of the 9th ACM International Workshop on Security and Privacy Analytics}, pages 27--32, 2023.

\bibitem[McLean et~al.(2023)McLean, Read, Thompson, Baber, Stanton, and Salmon]{mclean2023risks}
Scott McLean, Gemma~JM Read, Jason Thompson, Chris Baber, Neville~A Stanton, and Paul~M Salmon.
\newblock The risks associated with artificial general intelligence: A systematic review.
\newblock \emph{Journal of Experimental \& Theoretical Artificial Intelligence}, 35\penalty0 (5):\penalty0 649--663, 2023.

\bibitem[McMahon and Threats(2023)]{mcmahon2023maligned}
Dave McMahon and Pacing Threats.
\newblock Maligned influence and interference in canada.
\newblock \emph{CGAI: Canadian Global Affairs Institute}, 2023.

\bibitem[Meta(2023)]{metainauthentic}
Meta.
\newblock Inauthentic {Behavior} {\textbar} {Transparency} {Center}, 2023.
\newblock URL \url{https://transparency.meta.com/en-gb/policies/community-standards/inauthentic-behavior/}.

\bibitem[Microsoft(2024)]{microsoftdefensereport2024}
Microsoft.
\newblock Microsoft digital defense report 2024, 10 2024.
\newblock URL \url{https://cdn-dynmedia-1.microsoft.com/is/content/microsoftcorp/microsoft/final/en-us/microsoft-brand/documents/Microsoft\%20Digital\%20Defense\%20Report\%202024\%20\%281\%29.pdf}.

\bibitem[Moreau and Rourke(2024)]{deepfakewomen2024}
Shona Moreau and Chloe Rourke.
\newblock Fake porn causes real harm to women, 02 2024.
\newblock URL \url{https://policyoptions.irpp.org/magazines/february-2024/fake-porn-harm/}.

\bibitem[Moti et~al.(2024)Moti, Senol, Bostani, Borgesius, Moonsamy, Mathur, and Acar]{moti2024targeted}
Zahra Moti, Asuman Senol, Hamid Bostani, Frederik~Zuiderveen Borgesius, Veelasha Moonsamy, Arunesh Mathur, and Gunes Acar.
\newblock Targeted and troublesome: Tracking and advertising on children’s websites.
\newblock In \emph{2024 IEEE Symposium on Security and Privacy (SP)}, pages 1517--1535. IEEE, 2024.

\bibitem[MTAC(2023)]{mtac_2023nov8}
MTAC.
\newblock Protecting election 2024 from foreign malign influence: lessons learned help us anticipate the future, 11 2023.
\newblock URL \url{https://blogs.microsoft.com/wp-content/uploads/prod/sites/5/2023/11/MTAC-Report-2024-Election-Threat-Assessment-11082023-2-1.pdf}.

\bibitem[MTAC(2024{\natexlab{a}})]{mtac_2024apr17}
MTAC.
\newblock Nation-states engage in us-focused influence operations ahead of us presidential election, 04 2024{\natexlab{a}}.
\newblock URL \url{https://blogs.microsoft.com/wp-content/uploads/prod/sites/5/2024/04/MTAC-Report-Elections-Report-Nation-states-engage-in-US-focused-influence-operations-ahead-of-US-presidential-election-04172024.pdf}.

\bibitem[MTAC(2024{\natexlab{b}})]{mtac_2024aug9}
MTAC.
\newblock Iran steps into us election 2024 with cyber-enabled influence operations, 08 2024{\natexlab{b}}.
\newblock URL \url{https://cdn-dynmedia-1.microsoft.com/is/content/microsoftcorp/microsoft/final/en-us/microsoft-brand/documents/5bc57431-a7a9-49ad-944d-b93b7d35d0fc.pdf}.

\bibitem[MTAC(2024{\natexlab{c}})]{mtac_2024oct23}
MTAC.
\newblock Russia, iran, and china continue influence campaigns in final weeks before election day 2024, 10 2024{\natexlab{c}}.
\newblock URL \url{https://cdn-dynmedia-1.microsoft.com/is/content/microsoftcorp/microsoft/msc/documents/presentations/CSR/MTAC-Election-Report-5-on-Russian-Influence.pdf}.

\bibitem[MTAC(2024{\natexlab{d}})]{mtac_2024sept17}
MTAC.
\newblock Russia leverages cyber proxies and volga flood assets in expansive influence efforts, 09 2024{\natexlab{d}}.
\newblock URL \url{https://cdn-dynmedia-1.microsoft.com/is/content/microsoftcorp/microsoft/msc/documents/presentations/CSR/MTAC-Election-Report-4.pdf}.

\bibitem[Muehlhauser and Salamon(2013)]{muehlhauser2013intelligence}
Luke Muehlhauser and Anna Salamon.
\newblock Intelligence explosion: Evidence and import.
\newblock In \emph{Singularity hypotheses: A scientific and philosophical assessment}, pages 15--42. Springer, 2013.

\bibitem[M{\"u}ller and Bostrom(2016)]{muller2016future}
Vincent~C M{\"u}ller and Nick Bostrom.
\newblock Future progress in artificial intelligence: A survey of expert opinion.
\newblock \emph{Fundamental issues of artificial intelligence}, pages 555--572, 2016.

\bibitem[Najafabadi and Domanski(2018)]{najafabadi2018hacktivism}
Mahdi~M Najafabadi and Robert~J Domanski.
\newblock Hacktivism and distributed hashtag spoiling on twitter: Tales of the\# irantalks.
\newblock \emph{First Monday}, 2018.

\bibitem[Nakutavičiūtė(2023)]{proxynordvpn2023}
Jomilė Nakutavičiūtė.
\newblock Proxy vs vpn: What are the main differences?, Mar 2023.
\newblock URL \url{https://nordvpn.com/blog/vpn-vs-proxy/?srsltid=AfmBOop98yBTwR8vCQc7a-MbPy0iz0LgylkTk_-2wIXp_PUCmqEgbaIi}.

\bibitem[{NATO StratCom COE}(2018)]{NATOstratcom2018}
{NATO StratCom COE}.
\newblock The black market for social media manipulation, 11 2018.
\newblock URL \url{https://stratcomcoe.org/cuploads/pfiles/web_nato_report_-__the_black_market_of_malicious_use_of_social_media-1.pdf}.

\bibitem[Newman and Schwarz(2023)]{newman2023misinformed}
Eryn~J Newman and Norbert Schwarz.
\newblock Misinformed by images: How images influence perceptions of truth and what can be done about it.
\newblock \emph{Current Opinion in Psychology}, page 101778, 2023.

\bibitem[Omohundro(2018)]{omohundro2018basic}
Stephen~M Omohundro.
\newblock The basic ai drives.
\newblock In \emph{Artificial intelligence safety and security}, pages 47--55. Chapman and Hall/CRC, 2018.

\bibitem[OpenAI(2024{\natexlab{a}})]{OpenAIcharter}
OpenAI.
\newblock {OpenAI Charter}, 09 2024{\natexlab{a}}.
\newblock URL \url{https://openai.com/charter/}.

\bibitem[OpenAI(2024{\natexlab{b}})]{openai2024cyber}
OpenAI.
\newblock Influence and cyber operations: an update, 10 2024{\natexlab{b}}.
\newblock URL \url{https://cdn.openai.com/threat-intelligence-reports/influence-and-cyber-operations-an-update_October-2024.pdf}.

\bibitem[Park et~al.(2024)Park, Goldstein, O’Gara, Chen, and Hendrycks]{park2024ai}
Peter~S Park, Simon Goldstein, Aidan O’Gara, Michael Chen, and Dan Hendrycks.
\newblock Ai deception: A survey of examples, risks, and potential solutions.
\newblock \emph{Patterns}, 5\penalty0 (5), 2024.

\bibitem[Pathak et~al.(2023)Pathak, Spezzano, and Pera]{pathak2023understanding}
Royal Pathak, Francesca Spezzano, and Maria~Soledad Pera.
\newblock Understanding the contribution of recommendation algorithms on misinformation recommendation and misinformation dissemination on social networks.
\newblock \emph{ACM Transactions on the Web}, 17\penalty0 (4):\penalty0 1--26, 2023.

\bibitem[Pelrine et~al.(2023)Pelrine, Imouza, Thibault, Reksoprodjo, Gupta, Christoph, Godbout, and Rabbany]{pelrine2023towards}
Kellin Pelrine, Anne Imouza, Camille Thibault, Meilina Reksoprodjo, Caleb Gupta, Joel Christoph, Jean-Fran{\c{c}}ois Godbout, and Reihaneh Rabbany.
\newblock Towards reliable misinformation mitigation: Generalization, uncertainty, and gpt-4.
\newblock \emph{arXiv preprint arXiv:2305.14928}, 2023.

\bibitem[Pennycook et~al.(2021)Pennycook, Epstein, Mosleh, Arechar, Eckles, and Rand]{pennycook2021shifting}
Gordon Pennycook, Ziv Epstein, Mohsen Mosleh, Antonio~A Arechar, Dean Eckles, and David~G Rand.
\newblock Shifting attention to accuracy can reduce misinformation online.
\newblock \emph{Nature}, 592\penalty0 (7855):\penalty0 590--595, 2021.

\bibitem[{Perplexity AI}(2024)]{dailies_2024}
{Perplexity AI}.
\newblock Openai’s 5 steps to agi, 07 2024.
\newblock URL \url{https://www.perplexity.ai/page/openai-s-5-steps-to-agi-STzklF5SSQ6JOiBTaV.cfA}.

\bibitem[Priestman et~al.(2019)Priestman, Anstis, Sebire, Sridharan, and Sebire]{priestman2019phishing}
Ward Priestman, Tony Anstis, Isabel~G Sebire, Shankar Sridharan, and Neil~J Sebire.
\newblock Phishing in healthcare organisations: Threats, mitigation and approaches.
\newblock \emph{BMJ health \& care informatics}, 26\penalty0 (1), 2019.

\bibitem[Ramer(2024)]{bidenrobocall2024}
Holly Ramer.
\newblock Political consultant behind fake biden robocalls says he was trying to highlight a need for ai rules, 02 2024.
\newblock URL \url{https://apnews.com/article/ai-robocall-biden-new-hampshire-primary-2024-f94aa2d7f835ccc3cc254a90cd481a99}.

\bibitem[Reddit(2024)]{redditads2024}
Reddit.
\newblock Reddit ad types: Reddit for business, 10 2024.
\newblock URL \url{https://www.business.reddit.com/advertise/ad-types}.

\bibitem[Rheault and Musulan(2021)]{rheault2021efficient}
Ludovic Rheault and Andreea Musulan.
\newblock Efficient detection of online communities and social bot activity during electoral campaigns.
\newblock \emph{Journal of Information Technology \& Politics}, 18\penalty0 (3):\penalty0 324--337, 2021.

\bibitem[Rieder and Skop(2021)]{rieder2021fabrics}
Bernhard Rieder and Yarden Skop.
\newblock The fabrics of machine moderation: Studying the technical, normative, and organizational structure of perspective api.
\newblock \emph{Big Data \& Society}, 8\penalty0 (2):\penalty0 20539517211046181, 2021.

\bibitem[Rogers and Niederer(2020)]{rogers2020politics}
Richard Rogers and Sabine Niederer.
\newblock \emph{The politics of social media manipulation}.
\newblock Amsterdam University Press, 2020.

\bibitem[Rosenberg(2023)]{rosenberg2023manipulation}
Louis Rosenberg.
\newblock The manipulation problem: conversational ai as a threat to epistemic agency.
\newblock \emph{arXiv preprint arXiv:2306.11748}, 2023.

\bibitem[Roser(2023)]{owidaitimelines}
Max Roser.
\newblock Ai timelines: What do experts in artificial intelligence expect for the future?
\newblock \emph{Our World in Data}, 02 2023.
\newblock https://ourworldindata.org/ai-timelines.

\bibitem[Ross et~al.(2019)Ross, Pilz, Cabrera, Brachten, Neubaum, and Stieglitz]{ross2019social}
Bj{\"o}rn Ross, Laura Pilz, Benjamin Cabrera, Florian Brachten, German Neubaum, and Stefan Stieglitz.
\newblock Are social bots a real threat? an agent-based model of the spiral of silence to analyse the impact of manipulative actors in social networks.
\newblock \emph{European Journal of Information Systems}, 28\penalty0 (4):\penalty0 394--412, 2019.

\bibitem[Ruffo et~al.(2023)Ruffo, Semeraro, Giachanou, and Rosso]{ruffo2023studying}
Giancarlo Ruffo, Alfonso Semeraro, Anastasia Giachanou, and Paolo Rosso.
\newblock Studying fake news spreading, polarisation dynamics, and manipulation by bots: A tale of networks and language.
\newblock \emph{Computer science review}, 47:\penalty0 100531, 2023.

\bibitem[Russell(2022)]{russell2022human}
Stuart Russell.
\newblock Human-compatible artificial intelligence., 2022.

\bibitem[Ryan-Mosley(2024)]{MITtechreview}
Tate Ryan-Mosley.
\newblock How generative {AI} is boosting the spread of disinformation and propaganda, 2024.
\newblock URL \url{https://www.technologyreview.com/2023/10/04/1080801/generative-ai-boosting-disinformation-and-propaganda-freedom-house/}.

\bibitem[Sailio et~al.(2020)Sailio, Latvala, and Szanto]{sailio2020cyber}
Mirko Sailio, Outi-Marja Latvala, and Alexander Szanto.
\newblock Cyber threat actors for the factory of the future.
\newblock \emph{Applied Sciences}, 10\penalty0 (12):\penalty0 4334, 2020.

\bibitem[Salvi et~al.(2024)Salvi, Ribeiro, Gallotti, and West]{salvi2024conversational}
Francesco Salvi, Manoel~Horta Ribeiro, Riccardo Gallotti, and Robert West.
\newblock On the conversational persuasiveness of large language models: A randomized controlled trial.
\newblock \emph{arXiv preprint arXiv:2403.14380}, 2024.

\bibitem[Saxonhouse(2005)]{saxonhouse2005free}
Arlene~W Saxonhouse.
\newblock \emph{Free speech and democracy in ancient Athens}.
\newblock Cambridge University Press, 2005.

\bibitem[Schia and Gjesvik(2020)]{schia2020hacking}
Niels~Nagelhus Schia and Lars Gjesvik.
\newblock Hacking democracy: managing influence campaigns and disinformation in the digital age.
\newblock \emph{Journal of Cyber Policy}, 5\penalty0 (3):\penalty0 413--428, 2020.

\bibitem[{Select Committee on Intelligence}(2019)]{senaterussianreport}
{Select Committee on Intelligence}.
\newblock Russian active measures campaigns and interference in the 2016 u.s. election.
\newblock \url{https://www.intelligence.senate.gov/sites/default/files/documents/Report_Volume2.pdf}, 2019.

\bibitem[Seo(2021)]{seo2021amplifying}
Mihye Seo.
\newblock Amplifying panic and facilitating prevention: Multifaceted effects of traditional and social media use during the 2015 mers crisis in south korea.
\newblock \emph{Journalism \& Mass Communication Quarterly}, 98\penalty0 (1):\penalty0 221--240, 2021.

\bibitem[Simchon et~al.(2023)Simchon, Sutton, Edwards, and Lewandowsky]{simchon2023online}
Almog Simchon, Adam Sutton, Matthew Edwards, and Stephan Lewandowsky.
\newblock Online reading habits can reveal personality traits: towards detecting psychological microtargeting.
\newblock \emph{PNAS nexus}, 2\penalty0 (6):\penalty0 pgad191, 2023.

\bibitem[Sotala(2018)]{sotala2018disjunctive}
Kaj Sotala.
\newblock Disjunctive scenarios of catastrophic ai risk.
\newblock In \emph{Artificial intelligence safety and security}, pages 315--337. Chapman and Hall/CRC, 2018.

\bibitem[Spencer(2020)]{spencer2020problem}
Shaun~B Spencer.
\newblock The problem of online manipulation.
\newblock \emph{U. Ill. L. Rev.}, page 959, 2020.

\bibitem[Spring(2024)]{bbcconspiracyprofit}
Marianna Spring.
\newblock How hurricane milton and helene conspiracy theories took over social media, 10 2024.
\newblock URL \url{https://www.bbc.com/news/articles/c1e8q50y3v7o}.

\bibitem[Sun et~al.(2023)Sun, Jia, Hou, and Lyu]{sun2023ai}
Chengzhe Sun, Shan Jia, Shuwei Hou, and Siwei Lyu.
\newblock Ai-synthesized voice detection using neural vocoder artifacts.
\newblock In \emph{Proceedings of the IEEE/CVF Conference on Computer Vision and Pattern Recognition}, pages 904--912, 2023.

\bibitem[Susser et~al.(2019)Susser, Roessler, and Nissenbaum]{susser2019online}
Daniel Susser, Beate Roessler, and Helen Nissenbaum.
\newblock Online manipulation: Hidden influences in a digital world.
\newblock \emph{Geo. L. Tech. Rev.}, 4:\penalty0 1, 2019.

\bibitem[Tang and Wikström(2024)]{aislop2024}
Jiaru Tang and Patrik Wikström.
\newblock “side job, self-employed, high-paid”: Behind the ai slop flooding tiktok and facebook, 09 2024.
\newblock URL \url{https://theconversation.com/side-job-self-employed-high-paid-behind-the-ai-slop-flooding-tiktok-and-facebook-237638}.

\bibitem[Tian et~al.(2024)Tian, Yu, Orlovskiy, Vergho, Rivera, Goel, Yang, Godbout, Rabbany, and Pelrine]{tian2024web}
Jacob-Junqi Tian, Hao Yu, Yury Orlovskiy, Tyler Vergho, Mauricio Rivera, Mayank Goel, Zachary Yang, Jean-Francois Godbout, Reihaneh Rabbany, and Kellin Pelrine.
\newblock Web retrieval agents for evidence-based misinformation detection.
\newblock \emph{arXiv preprint arXiv:2409.00009}, 2024.

\bibitem[Tomassi et~al.(2024)Tomassi, Falegnami, and Romano]{tomassi2024mapping}
Andrea Tomassi, Andrea Falegnami, and Elpidio Romano.
\newblock Mapping automatic social media information disorder. the role of bots and ai in spreading misleading information in society.
\newblock \emph{Plos one}, 19\penalty0 (5):\penalty0 e0303183, 2024.

\bibitem[Tomz(2007)]{tomz2007domestic}
Michael Tomz.
\newblock Domestic audience costs in international relations: An experimental approach.
\newblock \emph{International organization}, 61\penalty0 (4):\penalty0 821--840, 2007.

\bibitem[Torkington(2024)]{wef2024}
Simon Torkington.
\newblock These are the 3 biggest emerging risks the world is facing, 01 2024.
\newblock URL \url{https://www.weforum.org/agenda/2024/01/ai-disinformation-global-risks/}.

\bibitem[Torres-Lugo et~al.(2022)Torres-Lugo, Yang, and Menczer]{torres2022manufacture}
Christopher Torres-Lugo, Kai-Cheng Yang, and Filippo Menczer.
\newblock The manufacture of partisan echo chambers by follow train abuse on twitter.
\newblock In \emph{Proceedings of the International AAAI Conference on Web and Social Media}, volume~16, pages 1017--1028, 2022.

\bibitem[Treen et~al.(2020)Treen, Williams, and O'Neill]{treen2020online}
Kathie M~d'I Treen, Hywel~TP Williams, and Saffron~J O'Neill.
\newblock Online misinformation about climate change.
\newblock \emph{Wiley Interdisciplinary Reviews: Climate Change}, 11\penalty0 (5):\penalty0 e665, 2020.

\bibitem[Tucker et~al.(2018)Tucker, Guess, Barber{\'a}, Vaccari, Siegel, Sanovich, Stukal, and Nyhan]{tucker2018social}
Joshua~A Tucker, Andrew Guess, Pablo Barber{\'a}, Cristian Vaccari, Alexandra Siegel, Sergey Sanovich, Denis Stukal, and Brendan Nyhan.
\newblock Social media, political polarization, and political disinformation: A review of the scientific literature.
\newblock \emph{Political polarization, and political disinformation: a review of the scientific literature (March 19, 2018)}, 2018.

\bibitem[{U.S. Department of Justice}(2024)]{doppelganger}
{U.S. Department of Justice}, 09 2024.
\newblock URL \url{https://www.justice.gov/d9/2024-09/doppelganger_affidavit_9.4.24.pdf}.

\bibitem[{US Department of State}(2024)]{usdeptstate}
{US Department of State}.
\newblock Election {Security}: {U}.{S}. {Government}’s {Efforts} to {Protect} the 2024 {U}.{S}. {Election} from {Foreign} {Malign} {Influence}.
\newblock Press interview, 09 2024.
\newblock URL \url{https://www.state.gov/briefings-foreign-press-centers/protecting-the-2024-election-from-foreign-malign-influence/}.

\bibitem[Vergho et~al.(2024)Vergho, Godbout, Rabbany, and Pelrine]{vergho2024comparing}
Tyler Vergho, Jean-Francois Godbout, Reihaneh Rabbany, and Kellin Pelrine.
\newblock Comparing gpt-4 and open-source language models in misinformation mitigation.
\newblock \emph{arXiv preprint arXiv:2401.06920}, 2024.

\bibitem[Vosoughi et~al.(2018)Vosoughi, Roy, and Aral]{vosoughi2018spread}
Soroush Vosoughi, Deb Roy, and Sinan Aral.
\newblock The spread of true and false news online.
\newblock \emph{science}, 359\penalty0 (6380):\penalty0 1146--1151, 2018.

\bibitem[vpattnaik et~al.(2024)vpattnaik, eavena, diannegali, denisebmsft, Dansimp, chrisda, and vpattnai]{microsoftthreatactors2024}
vpattnaik, eavena, diannegali, denisebmsft, Dansimp, chrisda, and vpattnai.
\newblock How microsoft names threat actors - microsoft defender xdr, 08 2024.
\newblock URL \url{https://learn.microsoft.com/en-us/defender-xdr/microsoft-threat-actor-naming}.

\bibitem[Vraga and Bode(2020)]{vraga2020defining}
Emily~K Vraga and Leticia Bode.
\newblock Defining misinformation and understanding its bounded nature: Using expertise and evidence for describing misinformation.
\newblock \emph{Political Communication}, 37\penalty0 (1):\penalty0 136--144, 2020.

\bibitem[Walter and Ophir(2023)]{walter2023trolls}
Dror Walter and Yotam Ophir.
\newblock Trolls without borders: a comparative analysis of six foreign countries’ online propaganda campaigns.
\newblock \emph{Human Communication Research}, 49\penalty0 (4):\penalty0 421--432, 2023.

\bibitem[Weld et~al.(2024)Weld, Zhang, and Althoff]{weld2024making}
Galen Weld, Amy~X Zhang, and Tim Althoff.
\newblock Making online communities ‘better’: a taxonomy of community values on reddit.
\newblock In \emph{Proceedings of the International AAAI Conference on Web and Social Media}, volume~18, pages 1611--1633, 2024.

\bibitem[Yang et~al.(2021)Yang, Pierri, Hui, Axelrod, Torres-Lugo, Bryden, and Menczer]{yang2021covid}
Kai-Cheng Yang, Francesco Pierri, Pik-Mai Hui, David Axelrod, Christopher Torres-Lugo, John Bryden, and Filippo Menczer.
\newblock The covid-19 infodemic: Twitter versus facebook.
\newblock \emph{Big Data \& Society}, 8\penalty0 (1):\penalty0 20539517211013861, 2021.

\bibitem[Yurtseven et~al.(2021)Yurtseven, Bagriyanik, and Ayvaz]{yurtseven2021review}
{\.I}lke Yurtseven, Selami Bagriyanik, and Serkan Ayvaz.
\newblock A review of spam detection in social media.
\newblock In \emph{2021 6th International Conference on Computer Science and Engineering (UBMK)}, pages 383--388. IEEE, 2021.

\bibitem[Zannettou et~al.(2019)Zannettou, Caulfield, Setzer, Sirivianos, Stringhini, and Blackburn]{zannettou2019let}
Savvas Zannettou, Tristan Caulfield, William Setzer, Michael Sirivianos, Gianluca Stringhini, and Jeremy Blackburn.
\newblock Who let the trolls out? towards understanding state-sponsored trolls.
\newblock In \emph{Proceedings of the 10th acm conference on web science}, pages 353--362, 2019.

\bibitem[Zhen et~al.(2023)Zhen, Yan, Tang, Nan, and Yang]{zhen2023social}
Lichen Zhen, Bei Yan, Jack~Lipei Tang, Yuanfeixue Nan, and Aimei Yang.
\newblock Social network dynamics, bots, and community-based online misinformation spread: lessons from anti-refugee and covid-19 misinformation cases.
\newblock \emph{The Information Society}, 39\penalty0 (1):\penalty0 17--34, 2023.

\end{thebibliography}

\end{document}